\documentclass[aps,prd,twocolumn,nofootinbib,superscriptaddress,amsmath,amssymb,floatfix]{revtex4}

\usepackage{amsfonts}
\usepackage{bbm}
\usepackage{microtype}
\usepackage{color,soul}
\usepackage{tikz,ragged2e}
\usepackage[latin9]{inputenc}
\usepackage{array}
\usepackage{booktabs}
\usepackage[justification=justified,singlelinecheck=false]{caption}
\usepackage{tabularx,booktabs,caption}
\usepackage{relsize}
\usepackage{graphicx}
\usepackage{dcolumn}
\usepackage{bm,color}
\usepackage{hhline}
\usepackage{multirow}

\usepackage[
singlelinecheck=false
]{caption}
\usepackage{blindtext}
\usepackage{amsthm}
\newtheorem{defn}{Definition}
\allowdisplaybreaks
\usepackage{hhline}
\usepackage{multirow}

\usepackage{scrextend}
\usepackage[
singlelinecheck=false
]{caption}
\usepackage{blindtext}
\newcommand{\specialcell}[2][c]{%
\begin{tabular}[#1]{@{}c@{}}#2\end{tabular}}
\usepackage{enumerate}
\makeatletter
\def\@footnotecolor{red}
\define@key{Hyp}{footnotecolor}{%
 \HyColor@HyperrefColor{#1}\@footnotecolor%
}
\def\@footnotemark{%
    \leavevmode
    \ifhmode\edef\@x@sf{\the\spacefactor}\nobreak\fi
    \stepcounter{Hfootnote}%
    \global\let\Hy@saved@currentHref\@currentHref
    \hyper@makecurrent{Hfootnote}%
    \global\let\Hy@footnote@currentHref\@currentHref
    \global\let\@currentHref\Hy@saved@currentHref
    \hyper@linkstart{footnote}{\Hy@footnote@currentHref}%
    \@makefnmark
    \hyper@linkend
    \ifhmode\spacefactor\@x@sf\fi
    \relax
}
\makeatother
\usepackage[colorlinks=true,  citecolor=red, urlcolor=black, linkcolor=blue, footnotecolor=blue]{hyperref}

\begin{document}

\title{The Maxwell-Proca theory: definition and construction}

\author{Ver\'onica Errasti D\'iez}
\email{ veroerdi@mppmu.mpg.de}
\affiliation{Max-Planck-Institut f\"ur Physik (Werner-Heisenberg-Institut)\\
 F\"ohringer Ring 6, 80805 Munich, Germany}
\author{Brage Gording}
\email{ brageg@mppmu.mpg.de}
\affiliation{Max-Planck-Institut f\"ur Physik (Werner-Heisenberg-Institut)\\
 F\"ohringer Ring 6, 80805 Munich, Germany}
\author{Julio A. M\'endez-Zavaleta}
\email{julioamz@mpp.mpg.de}
\affiliation{Max-Planck-Institut f\"ur Physik (Werner-Heisenberg-Institut)\\
 F\"ohringer Ring 6, 80805 Munich, Germany}
\author{Angnis Schmidt-May}
\email{angnissm@mpp.mpg.de}
\affiliation{Max-Planck-Institut f\"ur Physik (Werner-Heisenberg-Institut)\\
 F\"ohringer Ring 6, 80805 Munich, Germany}

\begin{abstract}
We present a systematic construction of the most general first order Lagrangian
describing an arbitrary number of interacting Maxwell and Proca fields on Minkowski spacetime.
To this aim, we first formalize the notion of a Proca field, in analogy to the well known Maxwell field.
Our definition allows for a non-linear realization of the Proca mass, in the form of derivative self-interactions.
Consequently, we consider so-called generalized Proca/vector Galileons.
We explicitly demonstrate the ghost-freedom of this complete Maxwell-Proca theory by obtaining its constraint algebra.
We find that, when multiple Proca fields are present, their interactions must fulfill non-trivial differential relations
in order to ensure the propagation of the correct number of degrees of freedom.
These relations had so far been overlooked, which means previous multi-Proca proposals generically contain ghosts.
This is a companion paper to arXiv:1905.06967 [hep-th]. It puts on a solid footing the theory there introduced.
\end{abstract}

\maketitle

\section{Introduction: a brief history of beyond electrodynamics \label{intro}}
In the second half of the 19th century, the foundations of classical electromagnetism were laid.
The Lorentz force law, together with Maxwell's equations,
provided a unified framework to explain the origin and propagation of electric and magnetic fields.
Light was then understood to be an electromagnetic phenomenon.
This body of work has since become one of the cornerstones of theoretical physics.
It is therefore not surprising that ongoing efforts are made so as to generalize Maxwell's theory.
In the following, we discuss the subset of extensions that provides the
relevant ingredients to our new proposal: the Maxwell-Proca theory.
(If the reader arrives from~\cite{ErrastiDiez:2019trb}, we now elaborate on the spin-one void noted in that introduction.)

\emph{Early extensions.}
As a field theory, Maxwell electrodynamics describes an Abelian massless vector field and its linear interactions with sources.
During the first half of the 20th century, extensions to this setup were introduced in order to formalize new observations.
For instance, consider nuclear phenomena.
As pointed out by Yukawa~\cite{Yukawa:1935xg}, the forces there implicated do not obey a Coulomb law,
but show a faster decay over distance.
This insight was in good agreement with the theory of a \emph{massive}
vector field introduced by Proca~\cite{Proca:1900nv}.
In his paper, Proca proposed a relativistic massive wave equation for a vector field
(analogous to the Klein-Gordon equation~\cite{KleinG}).
Subsequent work~\cite{Proca:1937} refined the implications of his proposal:
it was to be understood as a model of a massive spin-one force carrier, i.e.~a massive photon.
Of course, the phenomenologies of the Maxwell and Proca fields are different.
For example, the Proca field explicitly breaks the gauge symmetry and thus propagates one more degree of freedom than the Maxwell field.

Alternatively, this helicity-zero mode can be excited without spoiling gauge invariance.
Shortly after Proca's work, Bopp~\cite{Bopp:1940} and Podolsky~\cite{Podolsky:1942}
presented a model of a $U(1)$-invariant Abelian vector field with a modified kinetic structure.
Such novel electrodynamics (and generalizations~\cite{Horvath:1957}) was introduced to amend some of the theoretical concerns at the time.
Specifically, this was a consistent proposal describing short distance interactions and thus resolving the self-energy problem
---a neat review of which can be found in~\cite{Jo}.
Even though the proposed Lagrangians contain higher order {\it self-derivative} interactions,
the equations of motion remain second order\footnote{\label{Ostrfoot}Usually,
higher order field equations are avoided
by restricting to first order Lagrangians. This straightforwardly
prevents the undesired Ostrogradsky instability~\cite{Ostr}.}.
However, some other peculiar characteristics appear, such as non-locality.

Many other theories of relativistic vector fields emerged.
A prominent representative is the theory in which the
Maxwell kinetic term is promoted to an arbitrary smooth function of itself.
Concretely, an early proposal of this kind was Born-Infeld electrodynamics~\cite{BI}.
In this model, the gauge symmetry is preserved and the equations of motion are again second order.
Besides, Born-Infeld theory (and closely related constructions like the Heisenberg-Euler
model~\cite{Heisenberg:1935qt}) manages to resolve the infinities in the self-energy of the electron.
This is achieved at a price, though: the field equations become {\it non-linear}.
In fact, all theories sharing such properties are commonly labeled non-linear electrodynamics~\cite{Plebanski:1970}.
We stress that mass, derivative self-interactions and non-linearity are constituents of the Maxwell-Proca theory here constructed.

\emph{Modern proposals.}
In recent times, attempts have been made in pursuit of the most
general theory of a massive vector field whose self-interactions extend the standard Proca mass term.
Such theories are collectively known as \emph{generalized Proca} or
\emph{vector Galileon} ---a name that will be clarified later.
This picture was first suggested in~\cite{Tasinato:2014eka}.
The approach is based on the inclusion of derivative self-interactions,
carefully chosen so as to not excite further modes than the three expected in four spacetime dimensions.
Throughout the paper, we refer to any further degree of freedom carried by the action as a \emph{ghost}.
The motivation behind these models goes beyond extending previous electromagnetic-like theories:
when considered over a cosmological background, generalized Proca fields can give rise to self-acceleration
and fit data from late-time observations~\cite{DeFelice:2016yws}.
Evidently, a coupling to gravity is required for such applications.
Even though this coupling is already used in the initial references~\cite{Tasinato:2014eka},
a rigorous counting of degrees of freedom is not carried out there.
This issue has been further investigated~\cite{Hull:2015uwa,Heisenberg:2016eld},
but still much is to be done in this direction\footnote{There is a
wider collection of papers related to vector Galileons in the presence of gravity.
Their aim is not to prove ghost-freedom.
Instead, submodels are taken as a source to tackle specific goals,
such as finding black hole or star configurations~\cite{ref:gravity}.
While the models that deal with a single vector field and yield second order equations of motion are automatically ghost-free,
we highlight both the pertinence and non-triviality of the pending proof of ghost-freedom in the remaining proposals.}.

In closer detail, the above mentioned theories are defined by the Lagrangian
density of an Abelian spin-one field that lives on Minkowski spacetime.
Schematically, the Lagrangian can be split into two parts:
\begin{align}
\label{eq:GenProca1}
\mathcal{L}=\mathcal{L}_{(0)}+\widehat{\mathcal{L}}.
\end{align}
The first piece $\mathcal{L}_{(0)}$ is an arbitrary smooth function built
from the field  itself and its (dual) field strength.
As such, it generalizes both non-linear electrodynamics and the original Proca theory.
On the other hand, the second piece $\widehat{\mathcal{L}}$ admits a closed form and contains
self-interactions composed of contractions between the field and its derivatives.
Contrary to $\mathcal{L}_{(0)}$, the derivatives within $\widehat{\mathcal{L}}$ do not come as field strengths exclusively,
but they are such that the Lagrangian remains first order.
In fact, it has been the work of a couple of years to track down
the complete set of interactions comprising
the so-claimed most general $\widehat{\mathcal{L}}$~\cite{Tasinato:2014eka,Allys:2015sht,Allys:2016jaq}.
The interested reader can find thorough reviews on these theories in~\cite{Heisenberg:2017mzp}.

The generalized Proca theories introduced so far share a characteristic that has been exploited very actively.
When inspecting the longitudinal mode's dynamics in the decoupling limit,
one finds that it matches the acclaimed scalar Galileon.
Based on this feature, generalized Proca theories are interchangeably labeled vector Galileons.
First introduced in~\cite{Nicolis:2008in}, Galileon fields also arise in the context of modified gravity,
for instance in models that give an effective mass to the graviton~\cite{Trodden:2011xh}.
Both scalar and vector Galileons are close relatives of Horndeski's construction.
The latter gives the most general theory of a self-gravitating scalar~\cite{Horn} or vector (gauge) field~\cite{Horndeski:1976gi}.
Indeed, when Galileon-like theories are promoted to evolve in the presence of gravity,
they must couple to the curvature according to Horndeski's prescription.
Conversely, Galileons can be understood as the zero-curvature limit of a Horndeski theory.

Finally, theories of multiple spin-one fields have also been investigated,
albeit they have not received that much attention.
Interacting Maxwell fields were alluded to in the search of one-form Galileons in~\cite{multiMaxwell}.
In~\cite{Allys:2016kbq,Jimenez:2016upj}, theories that comprise several copies of generalized Proca fields were introduced.
There, a global rotational symmetry was imposed on the field space, which drastically reduced the allowed interactions.
The motivation behind this restriction was to provide an innate source capable of supporting isotropic cosmologies.
To our knowledge, there are no models incorporating couplings between massless and massive vector fields.

\emph{Summary of results.}
In this work, we focus on ghost-free interactions of multiple real Abelian fields on Minkowski spacetime.
Loosely speaking, these can be either massless (Maxwell) or massive (Proca).
We formalize both notions by providing their canonical definitions in terms of Hamiltonian constraints.
Then, we present a bottom-up construction of the most general first order Lagrangian involving such fields.
Our result, the Maxwell-Proca theory, is axiomatically complete.
It naturally contains Maxwell electromagnetism, non-linear electrodynamics and Proca's theory, together with its modern generalizations,
as particular subcases.
The general structure of the Maxwell-Proca interactions is similar to the generalized Proca action in (\ref{eq:GenProca1}),
but it is subject to highly restrictive differential conditions on the form of $\widehat{\mathcal{L}}$ when multiple Proca fields are involved.
The origin of these conditions lies at the secondary level of the Lagrangian constraint analysis, which we explain in detail.

\emph{Organization of the paper.}
In Sec.~\ref{sec:axiomas}, we introduce the mathematical formalism behind our setup.
All information here presented is based on the Lagrangian-Hamiltonian constraint analysis in Appendix \ref{app:GhostFreedom}.
In Sec.~\ref{sec:construction}, we systematically generate the exhaustive set of ghost-free interactions.
We continue in Sec.~\ref{sec:beyondMP}, discussing direct strategies to enhance the generality of the Maxwell-Proca theory.
At last, in Sec.~\ref{sec:remarks}, we summarize the key ideas behind the technical details of our construction.

\emph{Conventions.}
We work on $d$-dimensional Minkowski spacetime with the mostly positive metric signature $(-,+,\dots,+)$.
Spacetime indices are denoted by the Greek letters $(\mu,\nu,\dots)$ and raised/lowered with $\eta_{\mu\nu}=diag(-1,1,\dots,1)$ and its inverse $\eta^{\mu\nu}$.
Latin indices $(i,j,\dots)$ designate spatial coordinates and are trivially raised/lowered.
The alphabets $(\alpha, \beta,\dots)$ label massive fields, while their barred counterparts $(\bar{\alpha}, \bar{\beta},\dots)$ label massless fields.
Capital alphabets $(A,B,\dots)$ and $(A_1,A_2,\dots)$ enumerate Abelian fields, massive and massless alike.
All field labels are trivially raised/lowered.
Einstein summation convention is to be understood throughout.

\section{Axiomatization \label{sec:axiomas}}
The Maxwell and  Proca actions are the elemental theories of Abelian vector fields:
a massless one in the former case and a massive one in the latter.
In $d$-dimensional Minkowski spacetime and free of sources, the Lagrangian densities of these theories have the form
\begin{align}
\label{eq:MPLag}
\mathcal{L}_{\textrm{M}}=-\frac{1}{4}A_{\mu\nu}A^{\mu\nu},\qquad
\mathcal{L}_{\textrm{P}}=-\frac{1}{4}B_{\mu\nu}B^{\mu\nu}-\frac{1}{2}m^{2}B_{\mu}B^{\mu},
\end{align}
respectively.
We take both fields to be real.
Here, $A_{\mu\nu}=\partial_\mu A_\nu-\partial_\nu A_\mu$ is the field strength of the Maxwell field.
Similarly,  $B_{\mu\nu}$ is that of the Proca field and the constant  $m\neq0$ parameterizes its mass.
We begin by convening the essentials of these theories, since they serve as building blocks for the more elaborate Maxwell-Proca theory we propose.

Maxwell electrodynamics is manifestly invariant under the $U(1)$ gauge transformation of the second kind
\begin{align}
\label{eq:U1A}
A_{\mu}\longrightarrow A_{\mu}+\partial_{\mu} \Phi,
\end{align}
where $\Phi$ is an arbitrary real scalar field.
This means that $A_\mu$ transforms as a connection, in the affine representation of the gauge group.
We highlight that both the Lagrangian (off-shell) and the field equations (on-shell) are  invariant under (\ref{eq:U1A}).
The explicit mass term that tells apart both Lagrangians in (\ref{eq:MPLag}) has the crucial effect of
spoiling such gauge symmetry for the Proca field.

One of the central points throughout the paper is the count of the degrees of freedom propagated by a given Lagrangian,
as a means to ensure the absence of ghosts.
It is well known that  $\mathcal{L}_{\textrm{M}}$ propagates $d-2$ whereas $\mathcal{L}_{\textrm{P}}$
propagates $d-1$ degrees of freedom.
In four spacetime dimensions, Maxwell's theory accounts for two transversal polarization states,
while the Proca field incorporates an additional longitudinal mode.

Both $\mathcal{L}_{\textrm{M}}$ and $\mathcal{L}_{\textrm{P}}$ are singular.
In other words, the determinant of their \emph{primary Hessians},
\begin{align}
\label{eq:Hessian}
W_{\mu\nu}^{(\textrm{M})}:=\frac{\partial^{2}\mathcal{L}_{\textrm{M}}}{\partial\dot{A}^{\mu} \partial\dot{A}^{\nu}}\,, \qquad
W_{\mu\nu}^{(\textrm{P})}:=\frac{\partial^{2}\mathcal{L}_{\textrm{P}}}{\partial\dot{B}^{\mu} \partial\dot{B}^{\nu}}\,,
\end{align}
vanishes.
Here, $\dot{A}^{\mu}:=\partial_{0}A^{\mu}$ stands for the velocity of  the massless field
while $\dot{B}^{\mu}$ corresponds to that of the massive field.
As a consequence, some of the Euler-Lagrange equations following from (\ref{eq:MPLag}) are first order.
Namely, they are not equations of motion ---in the sense that they do not
involve accelerations of the fields--- but relations between the field variables and their velocities.
These relations are the primary Lagrangian constraints\footnote{This discussion applies to field theories with irreducible constraints.
All $\mathcal{L}_{\textrm{M}}$, $\mathcal{L}_{\textrm{P}}$ and the later introduced
Maxwell-Proca theory pertain to this class.
The case involving reducible constraints is briefly reviewed in Appendix \ref{app:LagCons}.}
of the system (if tautologically true, they can be further distinguished as identities).
They are responsible for the mismatch between the number of {\it a priori}
degrees of freedom and those that are physically meaningful.

Given any singular Lagrangian, the existence of primary constraints $\varphi=0$
that are not identically zero starts a renowned iterative procedure
to obtain the remaining constraints.
This algorithm is suitably referenced and self-consistently reviewed in Appendix \ref{app:LagCons}.
The gist of it is as follows.
The demand that the primary constraints be stable in time $\dot{\varphi}\overset{!}{=}0$
yields the so-called secondary Euler-Lagrange equations.
Those which are independent of accelerations, if any, are the secondary Lagrangian constraints $\phi=0$.
When extant, the stability of those secondary constraints that are not identically vanishing is not guaranteed.
Rather, requiring $\dot{\phi}\overset{!}{=}0$ leads to the third iteration.
The procedure terminates when no Lagrangian constraints appear or when they are all identically zero,
since then their stability need not be assured.

Going back to the Maxwell and Proca theories in (\ref{eq:MPLag}),
the just described constraint analysis tells us that
Maxwell electrodynamics is endowed with a single primary (Bianchi) identity,
while its massive counterpart contains two Lagrangian constraints: one primary and one secondary.
Furthermore, the generator of the gauge symmetry inherent to Maxwell's
theory can be regarded as an additional primary identity.

Knowledge of the identities and Lagrangian constraints in a
singular theory can be used to determine the degrees of freedom it propagates.
Most often, this is done by means of the Hamiltonian formula proposed by Dirac long ago:
\begin{align}
n_{\textrm{dof}}=M_0-N_1-\frac{N_2}{2},
\label{ndof0}
\end{align}
where $M_{0}$ is the number of field variables
and $(N_{1},N_2)$ denote the number of first and second class constraints, respectively.
Recall that a first class constraint is one that yields a vanishing
Poisson bracket with all of the constraints in the theory,
whereas a second class constraint does not.
It is worth noting that in~\cite{Diaz:2014yua} an alternative expression to
(\ref{ndof0}) was developed for particle systems,
which depends solely on Lagrangian quantities, including the off-shell gauge generators.
The result was later on adapted to field theories in~\cite{Diaz:2017tmy}.

Indeed, Appendix \ref{app:HamCons} is devoted to the exposition of a long
established systematic procedure to calculate $N_1$ and $N_2$ from the Lagrangian constraints.
As is widely known, the Bianchi and gauge identities in Maxwell's theory generate two first class constraints.
On the Proca side, the Lagrangian constraints translate to two second class constraints.
A quick inspection of (\ref{ndof0})  then yields the anticipated number of degrees of freedom.

Considering all the above exposed, we introduce the following two definitions:
\begin{defn}
\label{def:Maxwell}
A Maxwell field is any real Abelian vector field whose Lagrangian is associated with two first class constraints.
\end{defn}
\begin{defn}
\label{def:Proca}
A Proca field is any real Abelian vector field whose Lagrangian is associated with two second class constraints.
\end{defn}
From this point of view, the distinction between a Proca field~\cite{Proca:1900nv,Proca:1937}
and a generalized Proca field~\cite{Tasinato:2014eka,Allys:2015sht,Allys:2016jaq} is insubstantial.
Subsequently, we do not discern between the two.

Notice that, in the above definitions, the Poisson bracket is the relevant operation:
it distinguishes between first and second class constraints and thus defines the constraint algebra.
Bilinearity of the Poisson bracket implies linearity of the constraint algebra.
Consequently, a direct sum of whatever number of copies of the Maxwell and Proca Lagrangians
will preserve the constraint patterns in Definitions \ref{def:Maxwell} and \ref{def:Proca}.
This is generically not true when there is an explicit interaction involving (some of) the Proca fields.
Specifically, we prove in Appendix \ref{app:LagCons} that coupling massive vectors
alters the stability of their primary Lagrangian constraints.
What happens is that, at the secondary level, the constraints $\phi=0$ are absent.
Instead, we get independent Euler-Lagrange equations responsible for
exciting unwanted ghost-like degrees of freedom.

One of our key results lies in showing that one can always find befitting interactions
that prevent the aforementioned deviation from the Proca constraint scheme.
This is uniquely achieved by setting the \emph{secondary Hessian} to zero.
In more detail, consider some Lagrangian density $\mathcal{L}^{\prime}$ encoding the interaction of a set of $M$ Proca fields $\{B_{\mu}\}$.
The antisymmetric matrix
\begin{align}
\label{eq:secHess}
\widetilde{W}_{\alpha\beta}:=\frac{\partial^2\mathcal{L}^{\prime}}{\partial \dot{B}^{(\alpha)}_0\partial B^{(\beta)}_0}
-\frac{\partial^2 \mathcal{L}^{\prime}}{\partial \dot{B}^{(\beta)}_0\partial B^{(\alpha)}_0},
\end{align}
conforms the non-vanishing part of the secondary Hessian. Enforcing
\begin{align}
\label{eq:secondCER0}
\widetilde{W}_{\alpha\beta}\stackrel{!}{=}0\qquad\forall \alpha,\beta=1,\dots,M,
\end{align}
induces the appearance of the secondary constraints that restrain the ghost.
We stress that only when (\ref{eq:secondCER0}) is fulfilled,
can the vectors $\{B_{\mu}\}$ composing $\mathcal{L}^{\prime}$ be identified as Proca fields in the sense of Definition \ref{def:Proca}.
This prescription has been irreconcilably overlooked in the existing literature~\cite{Allys:2016kbq,Jimenez:2016upj}.
As a consequence, the models there submitted are ghost-full.

The preceding observations motivate the proposition of our main result:
\begin{defn}
\label{def:MaxPro}
The Maxwell-Proca theory is the most general Lagrangian of $N$ number of Maxwell fields and $M$ number of Proca fields,
such that
\begin{enumerate}[i]
\item The theory is defined over four-dimensional Minkowski spacetime.
\item The Lagrangian is at most first order.
\end{enumerate}
\end{defn}
By a first order Lagrangian density we mean
\begin{align}
\mathcal{L}=\mathcal{L}(A_{\mu},B_{\mu},\partial_\mu A_{\nu}, \partial_\mu B_{\nu}),
\end{align}
up to boundary terms.
Then, the variational principle will produce at most second order field equations.
This is a sufficient condition to safeguard our theory from Ostrogradsky-like instabilities, as already noted in footnote \ref{Ostrfoot}.
When a single Proca field is considered (i.e.~$M=1$), this is the necessary and sufficient condition.

The most concise form of the Maxwell-Proca theory is
\begin{align}
\label{eq:MPLagSchem}
\mathcal{L}_{\textrm{MP}}=\mathcal{L}_{\text{kin}}+\mathcal{L}_{\text{int}},
\end{align}
where the kinetic part can always be expressed as
\begin{align}
\label{eq:MPLkinSchem}
\mathcal{L}_{\text{kin}}=-\frac{1}{4}A_{\mu\nu}^{(\bar{\alpha})}A_{(\bar{\alpha})}^ {\mu\nu}-\frac{1}{4}B_{\mu\nu}^{({\alpha})}B_{({\alpha})}^ {\mu\nu},
\end{align}
and the interaction part encompasses every possible Lorentz scalar built out of the fields and their first derivatives,
while respecting the defining axioms.
We identify three sectors: massless-massless, massless-massive and the massive-massive.
Explicitly,
\begin{align}
\label{eq:MPLintSchem}
\mathcal{L}_{\text{int}}&=\mathcal{L}^{(AA)}+\mathcal{L}^{(AB)}+\mathcal{L}^{(BB)}.
\end{align}
We reassert that the Maxwell-Proca theory, in its final form of Sec.~\ref{sec:MassEigenbasis}, is only ghost-free when regarded as a whole
together with the condition that its secondary Hessian vanishes.
Whilst the massless interactions $\mathcal{L}^{(AA)}$ trivially fulfill this condition, $\mathcal{L}^{(AB)}$ and $\mathcal{L}^{(BB)}$ don't:
they are severely restricted by it.

\section{Construction of the ghost-free interactions \label{sec:construction}}
In this section, we systematically construct the interactions (\ref{eq:MPLintSchem}) building on the definitions given in the previous section.
We first review the necessary and sufficient conditions to ensure the existence of the suitable constraints at the primary level.
Albeit slightly more involved, this construction is very similar to that for a single Proca field~\cite{Tasinato:2014eka}.
It relies on the antisymmetric structures first recognized in~\cite{multiMaxwell} as a necessary and exhaustive tool
for developing healthy Lorentz invariant theories.
After implementing consistency at the primary level,
we present the highly non-trivial conditions that arise at the secondary level of the constraint algebra.
We discuss how these conditions notably restrict multi-Proca interactions and apply them to generate ghost-free interactions in several examples.

\subsection{Primary constraints\label{sec:PrimLevel}}
For our present purposes, it is useful to introduce a collective notation for all Abelian vector fields $X^{(A)}_\mu$,
where $A$ ranges from 1 to $M+N$.
This means that, for the moment, we set aside the issue of gauge invariance.

As already stated, healthiness of the Maxwell-Proca theory necessarily relies on a suitable rank reduction of its primary Hessian.
This is uniquely\footnote{For a first order theory, the rank of the Hessian can only be reduced by setting to zero entire rows.
Requiring linear combinations of rows to vanish amounts to a field redefinition.} achieved by
 \begin{align}
\label{eq:primCER}
\frac{\partial^2 \mathcal{L}_{\textrm{MP}}}{\partial \dot{X}^{(A)}_0\partial \dot{X}^{(B)}_\mu} \stackrel{!}{=} 0,\qquad\forall A,B.
\end{align}
Notice that (\ref{eq:primCER}) must hold for all $\mu$.
This is true also for the single Proca limit ($N=0$ and $M=1$),
as described, for instance, at the very beginning of page three in the review paper~\cite{Allys:2016jaq}.
We refer to the above as the \emph{primary constraint enforcing relations}, which are not to be confused  with the primary constraints themselves.
Together with the demand for Lorentz invariance, they will serve as a guiding principle to forge the interactions.
They are explicitly derived in the appendix.
Specifically, they are equivalent to (\ref{primcons}).

Next, recall that the Maxwell-Proca theory is of the form (\ref{eq:MPLagSchem}).
The kinetic part $\mathcal{L}_{\text{kin}}$ given in (\ref{eq:MPLkinSchem}) automatically satisfies the relations in (\ref{eq:primCER}).
We thus focus on the interactions $\mathcal{L}_{\text{int}}$.
In order to fulfill (\ref{eq:primCER}), they can have at most a linear dependence
on all the velocities associated with temporal components $\dot{X}^{(A)}_0$.
It is then natural to split $\mathcal{L}_{\text{int}}$ into terms $\mathcal{L}_{(0)}$ which do not depend on $\dot{X}^{(A)}_0$
and terms $\widehat{\mathcal{L}}_{\text{int}}$ which carry a linear dependence on these velocities:
\begin{align}
\label{eq:Lint}
\mathcal{L}_{\text{int}} = \mathcal{L}_{(0)} + \widehat{\mathcal{L}}_{\text{int}}.
\end{align}
By definition, $\mathcal{L}_{(0)}$ is an arbitrary real smooth function of the fields and their field strengths:
\begin{align}
\label{eq:L0piece}
\mathcal{L}_{(0)}=\mathcal{L}_{(0)}(X^{(A)}_{\mu}, X^{(A)}_{\mu\nu}).
\end{align}
Due to these loose restrictions, it contains infinitely many terms.

We now turn to the composition of $\widehat{\mathcal{L}}_{\text{int}}$.
For a first order Lagrangian and under the assumptions of smoothness and reality, $\widehat{\mathcal{L}}_{\text{int}}$ can be formally expanded as
\begin{align}\label{eq:LintFull}
\widehat{\mathcal{L}}_{\text{int}}= \sum_{n=1} ^\infty\mathcal{L}_{(n)},
\end{align}
with
\begin{align}
\label{eq:schemeLn}
\mathcal{L}_{(n)} = \mathcal{T}_{A_1\hdots A_n}^{\nu_1\hdots\nu_{n}\rho_1\hdots\rho_n}
\partial_{\nu_1} X^{(A_1)}_{\rho_1}\hdots\partial_{\nu_n} X^{(A_n)}_{\rho_n}.
\end{align}
Here, $n$ counts the number of derivatives and $(A_1,A_2,\dots)$ label all Abelian fields, as $(A,B,\dots)$ before.
The expansion coefficients $\mathcal{T}^{\nu_1...\rho_{n}}_{A_1...A_{n}}$ have an arbitrary dependence on the fields but contain no derivatives.
Note that, if all derivatives appear only in field strengths, the corresponding term belongs in $\mathcal{L}_{(0)}$ by definition.
Observe that the exchange of any triple $(A_i,\nu_i,\rho_i)\longleftrightarrow(A_j,\nu_j,\rho_j)$ leaves (\ref{eq:schemeLn}) invariant.
Hence, without loss of generality, we take the expansion coefficients to be symmetric under this replacement.

In order for $\widehat{\mathcal{L}}_{\text{int}}$ to satisfy (\ref{eq:primCER}), it is necessary and sufficient that
\begin{align}
\label{eq:FDPrimCER}
\frac{\partial{\widehat{\mathcal{L}}}_{\text{int}}}{\partial \dot{X}^{(A)}_0}
= \sum_{n=1}^\infty\frac{\partial \mathcal{L}_{(n)}}{\partial \dot{X}^{(A)}_{0}}
\end{align}
does not contain any time derivatives.
For $\mathcal{L}_{(1)}$ this is automatically true, while for
\begin{align}
\frac{\partial \mathcal{L}_{(n)}}{\partial \dot{X}^{(A)}_{0}}=
n\mathcal{T}_{AA_2\dots A_n}^{0\nu_2\dots\nu_n0\rho_2\dots\rho_n}
\partial_{\nu_2} X^{(A_2)}_{\rho_2}\!\!\dots\partial_{\nu_n} X^{(A_n)}_{\rho_n}
\end{align}
with $n>1$, it must be enforced.
The only Lorentz covariant way to do so is to entertain three distinct antisymmetrizations in (\ref{eq:schemeLn}):
\begin{enumerate}[i]
\item Antisymmetrization of all $n$ derivatives with each other. \label{ant1}
\item Antisymmetrization of all but one derivatives with each other and with all vector field indices.
Then, the respective $n-1$ fields appear only inside field strengths, while the derivative on the $n$th field
can be contracted in an arbitrary manner. \label{ant2}
\item Antisymmetrization of $k<n$ derivatives with $l<n$ fields, together with the
antisymmetrization of the remaining $n-k$ derivatives with the left-over $n-l$ fields. \label{ant3}
\end{enumerate}
All these options restrict but do not fully determine the coefficients $\mathcal{T}^{\nu_1...\rho_{n}}_{A_1...A_{n}}$.
Hence, the interactions contain free tensorial functions.

On four-dimensional Minkowski spacetime, the antisymmetrizations are uniquely accomplished using
the totally antisymmetric Levi-Civita tensor $\varepsilon_{\mu\nu\rho\sigma}$.
The above considerations show that all $\mathcal{L}_{(n>1)}$ must contain at least one Levi-Civita tensor.
There is another rather drastic consequence:
since we always need to antisymmetrize at least $n$ indices with each other,
the in principle infinite series expansion
(\ref{eq:LintFull}) truncates at $n=4$ in four spacetime dimensions.
Thus, the full interaction Lagrangian reads
\begin{align}\label{eq:fulllag}
\mathcal{L}_{\text{int}} = \mathcal{L}_{(0)}+\sum_{n=1}^4 \mathcal{T}_{A_1\hdots A_n}^{\nu_1\hdots\nu_{n}\rho_1\hdots\rho_n}
\partial_{\nu_1} X^{(A_1)}_{\rho_1}\hdots\partial_{\nu_n} X^{(A_n)}_{\rho_n}.
\end{align}
For concreteness, we explicitly write out all independent Lorentz contractions
that respect the primary constraint enforcing relation.

As stated earlier, for $\mathcal{L}_{(1)}$, there is no condition on the expansion coefficients:
\begin{align}
\mathcal{T}_{A}^{\nu\rho}
=\mathcal{F}_{A}{}^{\nu\rho}, \label{eq:T1}
\end{align}
where henceforth the spacetime tensors $\{\mathcal{F}\}$ are arbitrary smooth real functions\footnote{They need not be tensors with respect to the field
labels. For example, $\mathcal{F}_{A}{}^{\nu\rho}=\eta^{\nu\rho}$ for all $A$ is a valid choice for (\ref{eq:T1}).}.
Generically, they do not possess any symmetry properties and they are built from
the fields $X^{(A)}_{\mu}$ and the Lorentz invariant objects $\eta^{\mu\nu}$ and
$\varepsilon^{\mu\nu\rho\sigma}$.
The allowed coefficients in $\mathcal{L}_{(2)}$ are of the form
\begin{align}
\nonumber\mathcal{T}_{A_1A_2}^{\nu_1\nu_2\rho_1\rho_2}
=&\:\varepsilon^{\nu_1\nu_2}{}_{\mu_1\mu_2}\,\mathcal{F}_{A_1A_2}{}^{\mu_1\mu_2\rho_1\rho_2}\nonumber
\\ \nonumber
&{}+\varepsilon^{\rho_1\rho_2\nu_2}{}_{\mu}\,\mathcal{F}^{(1)}_{A_1A_2}{}^{\mu\nu_1}
\\
&{}+\varepsilon^{\nu_1\rho_2}{}_{\mu_1\mu_2}\varepsilon^{\nu_2\rho_1}{}_{\sigma_1\sigma_2}\,\mathcal{F}^{(2)}_{A_1A_2}{}^{\mu_1\mu_2\sigma_1\sigma_2}, \label{eq:T2}
\end{align}
where each line corresponds to the cases \ref{ant1}-\ref{ant3}, respectively.
For $\mathcal{L}_{(3)}$ we have
\begin{align}
\nonumber\mathcal{T}_{A_1A_2A_3}^{\nu_1\nu_2\nu_3\rho_1\rho_2\rho_3}
=&\:\varepsilon^{\nu_1\nu_2\nu_3}{}_{\mu}\,\mathcal{F}_{A_1A_2A_3}{}^{\mu\rho_1\rho_2\rho_3}
\nonumber\\ \nonumber
&+\varepsilon^{\nu_1\nu_3}{}_{\mu_1\mu_2}\,
\varepsilon^{\nu_2\rho_1\rho_2\rho_3}\,\mathcal{F}_{A_1A_2A_3}^{(1)}{}^{\mu_1\mu_2}
\\ \nonumber
&+\varepsilon^{\nu_1\nu_3\rho_1}{}_{\mu}\,
\varepsilon^{\nu_2\rho_2\rho_3}{}_{\sigma}\,\mathcal{F}_{A_1A_2A_3}^{(2)}{}^{\mu\sigma}
\\ \nonumber
&+\varepsilon^{\nu_1\nu_3\rho_2}{}_{\mu}\,
\varepsilon^{\nu_2\rho_1\rho_3}{}_{\sigma}\,\mathcal{F}_{A_1A_2A_3}^{(3)}{}^{\mu\sigma}
\\ \label{eq:T3}
&+\varepsilon^{\nu_1\nu_3\rho_1\rho_2}
\varepsilon^{\nu_2\rho_3}{}_{\sigma_1\sigma_2}\,\mathcal{F}_{A_1A_2A_3}^{(4)}{}^{\sigma_1\sigma_2}.
\end{align}
Here, the first term implements case \ref{ant1}. Case \ref{ant2} cannot be realized for $n>2$ in four dimensions.
The remaining terms all belong to and exhaust case \ref{ant3}.
Finally, in $\mathcal{L}_{(4)}$,
\begin{align}
\nonumber\mathcal{T}_{A_1A_2A_3A_4}^{\nu_1\nu_2\nu_3\nu_4\rho_1\rho_2\rho_3\rho_4}
=&\:\varepsilon^{\nu_1\nu_2\nu_3\nu_4}\,\mathcal{F}_{A_1A_2A_3A_4}{}^{\rho_1\rho_2\rho_3\rho_4}
\nonumber\\ \nonumber
{}&+\varepsilon^{\nu_1\nu_3\nu_4\rho_1}
\varepsilon^{\nu_2\rho_2\rho_3\rho_4}\,\mathcal{F}_{A_1A_2A_3A_4}^{(1)}
\\ \nonumber
{}&+\varepsilon^{\nu_1\nu_3\nu_4\rho_2}
\varepsilon^{\nu_2\rho_1\rho_3\rho_4}\,\mathcal{F}_{A_1A_2A_3A_4}^{(2)}
\\ \nonumber
{}&+\varepsilon^{\nu_1\nu_3\rho_1\rho_2}
\varepsilon^{\nu_2\nu_4\rho_3\rho_4}\,\mathcal{F}_{A_1A_2A_3A_4}^{(3)}
\\
{}&+\varepsilon^{\nu_1\nu_3\rho_2\rho_4}
\varepsilon^{\nu_2\nu_4\rho_1\rho_3}\,\mathcal{F}_{A_1A_2A_3A_4}^{(4)}, \label{eq:T4}
\end{align}
where the first term is associated to case \ref{ant1} while the remaining lines fully cover case \ref{ant3}.

To summarize, (\ref{eq:fulllag})-(\ref{eq:T4}) define the complete set of interactions that satisfy the primary constraint enforcing relations.

\subsection{Secondary constraints \label{sec:SecLevel}}
Our most crucial result, arising from the constraint analysis outlined in Appendix \ref{app:LagCons},
is the unveiling  of a set of \emph{secondary constraint enforcing relations}.
Namely, the axiomatically consistent closure of the constraint algebra requires the vanishing of the secondary Hessian
\begin{align}
\label{eq:secondCERX}
\widetilde{W}_{AB}:=\frac{\partial^2 \mathcal{L}_{\text{int}}}{\partial \dot{X}^{(A)}_0\partial X^{(B)}_0}
-\frac{\partial^2 \mathcal{L}_{\text{int}}}{\partial \dot{X}^{(B)}_0\partial X^{(A)}_0}\stackrel{!}{=} 0
\qquad \forall A,B,
\end{align}
with $\mathcal{L}_{\textrm{int}}$ as in (\ref{eq:Lint}).
Together with (\ref{eq:primCER}), the above gives the necessary and sufficient conditions to avoid the propagation of ghosts.
Then, the algebra closes automatically at the tertiary level.

The differential relations (\ref{eq:secondCERX}) further restrict the internal structure of the interactions constructed in the previous section,
which will be shortly exemplified.

\subsection{Gauge invariance\label{sec:MassEigenbasis}}
We now identify each Abelian vector $X^{(A)}$ with either a Maxwell or a Proca field.
Then, the interactions (\ref{eq:fulllag})-(\ref{eq:T4}) can be classified as in (\ref{eq:MPLintSchem}):
$\mathcal{L}^{(AA)}$ describing purely massless interactions,
$\mathcal{L}^{(BB)}$ for purely massive ones and
$\mathcal{L}^{(AB)}$ containing the interactions  between massless and massive fields.
In doing so, we need to respect gauge invariance.

All $N$ Maxwell fields ${A}^{(\bar{\alpha})}_{\mu}$ transform as connections under their own $U(1)$ gauge group, as in (\ref{eq:U1A}).
It has long been established that consequently the Maxwell vectors can only show up inside their field strengths.
As a result, the purely massless interactions are a subset of the general zeroth interactions:
$\mathcal{L}^{(AA)}\subseteq \mathcal{L}_{(0)}$.

The $M$ Proca fields $B^{({\alpha})}_{\mu}$, on the other hand, are gauge invariant by definition.
Therefore, the purely massive sector $\mathcal{L}^{(BB)}$ is straightforwardly obtained by replacing the fields $X^{(A)}$ with Proca fields:
\begin{align}\label{eq:LBB}
\mathcal{L}^{(BB)}=\sum_{n=0}^{4}\mathcal{L}^{(BB)}_{(n)},
\quad~~\mathcal{L}^{(BB)}_{(n)}:=\mathcal{L}_{(n)}\Big|_{X^{(A)}\rightarrow B^{(\alpha)}}.
\end{align}

The coupling $\mathcal{L}^{(AB)}$ between massless and massive fields is obtained in a similar way:
replacing $X^{(A)}$ with either ${A}^{(\bar{\alpha})}$ or $B^{(\alpha)}$,
with the Maxwell fields appearing only inside field strengths.
In particular, this means that the functions $\mathcal{F}$ depend exclusively on the Proca fields.

In terms of Maxwell and Proca fields, the secondary Hessian takes the form
\begin{align}
\widetilde{W}_{AB}=\begin{pmatrix}
\widetilde{W}_{\bar{\alpha}\bar{\beta}} &\widetilde{W}_{\bar{\alpha}{\beta}}\\
\widetilde{W}_{{\alpha}\bar{\beta}} & \widetilde{W}_{{\alpha}{\beta}}
\end{pmatrix}
\equiv
\begin{pmatrix}
0 & 0\\
0 & \widetilde{W}_{{\alpha}{\beta}}
\end{pmatrix}\,,
\end{align}
since Maxwell fields, which appear only inside field strengths, automatically satisfy (\ref{eq:secondCERX}).
For the same reason, the interactions $\mathcal{L}_{(0)}$ are not restricted at the secondary level.
In other words, (\ref{eq:secondCERX}) solely constrains the $n>0$ terms in $\mathcal{L}^{(AB)}$ and $\mathcal{L}^{(BB)}$ by
\begin{align}
\label{eq:secondCER}
\frac{\partial^2 \widehat{\mathcal{L}}_{\text{int}}}{\partial \dot{B}^{(\alpha)}_0\partial B^{(\beta)}_0}
-\frac{\partial^2 \widehat{\mathcal{L}}_{\text{int}}}{\partial \dot{B}^{(\beta)}_0\partial B^{(\alpha)}_0}\stackrel{!}{=} 0\qquad \forall \alpha,\beta.
\end{align}
Notice that only when two or more Proca fields are coupled, does (\ref{eq:secondCER}) become non-trivial.
The derivation of the above relations can be found in the appendix.
Concretely, (\ref{eq:secondCER}) is a rewriting of (\ref{mainres}).
The nature and relevance of these relations can be grasped as follows.
Equation (\ref{eq:primCER}) enforces the existence of a set of suitable (primary) constraints.
Then, equation (\ref{eq:secondCER}) guarantees their stability.
The (secondary) constraints that follow from equation (\ref{eq:secondCER}) hold true at all times, i.e.~they are automatically stable.
Therefore, no further constraint enforcing relations apply.

The internal structure of the expansion coefficients in (\ref{eq:T1})-(\ref{eq:T4}) is so rich
that implementing (\ref{eq:secondCER}) in full generality seems intractable,
because these conditions form a set of coupled non-linear partial differential equations.
Hence, we do not incorporate these relations into a building principle like we did at the primary level.
Instead, we provide a few examples of how to use them to extract ghost-free interactions.

\subsection{Explicit examples \label{sec:Examples}}
First, consider $\mathcal{L}^{(BB)}_{(1)}$, which remained unconstrained at the primary level.
It can be written in the form
\begin{align} \label{eq:L1BB}
\mathcal{L}^{(BB)}_{(1)}=\mathcal{F}_{\alpha}{}^{\mu\nu}\partial_{\mu}B^{(\alpha)}_{\nu},
\end{align}
where the coefficient is an arbitrary function of the Proca fields obeying $\mathcal{F}_{\alpha}{}^{\mu\nu}=\mathcal{F}_{\alpha}{}^{\nu\mu}$,
so as to exclude terms that belong in $\mathcal{L}^{(BB)}_{(0)}$.
Even this simple interaction is not guaranteed to be free of ghosts in a multi-Proca setup (i.e.~$M>1$).
The constraint enforcing relations (\ref{eq:secondCER}) reduce to
\begin{align} \label{eq:CER2L1BB}
\frac{ \partial \mathcal{F}_{\alpha}{}^{00}}{\partial B^{(\beta)}_{0}}-\frac{ \partial \mathcal{F}_{\beta}{}^{00}}{\partial B^{(\alpha)}_{0}}\overset{!}{=}0 \qquad \forall\alpha,\beta.
\end{align}
There are two straightforward solutions to this equation.
The functions $\mathcal{F}_{\alpha}{}^{\mu\nu}$ can be constants or they can depend exclusively on the field with label $\alpha$:
$\mathcal{F}_{\alpha}{}^{\mu\nu}=\mathcal{F}_{\alpha}{}^{\mu\nu}\left(B^{(\alpha)}\right)$
for all $\alpha$.
In the first case, the interaction becomes a boundary term.
In the second case, we are left with $M$ decoupled Proca theories.
While it is difficult to find the most general form of ${\mathcal{F}_{\alpha}}^{\mu\nu}$, one can always construct particular solutions.
A non-trivial solution fulfilling (\ref{eq:CER2L1BB}) is given in~\cite{ErrastiDiez:2019trb}.

Next, we take an example from~\cite{Allys:2016kbq}.
In our notation, their term reads
\begin{align}
\label{eq:L2BBex2}
\mathcal{L}^{(BB)}_{(2\text{-i})}
=& \big(
\varepsilon^{\rho_1\rho_2\mu_1\mu_2}\eta^{\sigma_1\sigma_2}
+ \varepsilon^{\sigma_1\sigma_2\mu_1\mu_2}\eta^{\rho_1\rho_2}\big)
B^{(\alpha)}_{\sigma_1}B^{(\beta)}_{\sigma_2} \nonumber\\
&\times  \varepsilon_{\mu_1\mu_2}{}^{\mu_3\mu_4}\,
\partial_{\mu_3}B_{\rho_1}^{(\alpha)}\partial_{\mu_4}B_{\rho_2}^{(\beta)}
\end{align}
and belongs in $\mathcal{L}^{(BB)}_{(2)}$.
For this interaction (which passes the primary level by completely antisymmetrizing the derivatives),
the secondary level (\ref{eq:secondCER}) requires
\begin{align}
\partial^i\left(B_0^{[(\alpha)} B_i^{(\beta)]}\right) \overset{!}{=} 0 \qquad \forall \alpha,\beta.
\end{align}
Clearly, this condition is not satisfied for generic field configurations.
The interaction term spoils the closure of the constraint algebra
and propagates additional degrees of freedom. We conclude that (\ref{eq:L2BBex2})
and various other interactions proposed in \cite{Allys:2016kbq,Jimenez:2016upj} carry ghosts.

We finish by providing a few examples of ghost-free interactions.
Any Lagrangian compatible with the primary constraint enforcing relations (\ref{eq:primCER}) that contains a single Proca field ($M=1$)
immediately passes the secondary level as well.
This is true for any number $N$ of Maxwell fields.
Accordingly, a large number of healthy
terms can directly be obtained from (\ref{eq:fulllag})-(\ref{eq:T4}).
A purely multi-Proca example ($M>1, N=0$) is
\begin{align}
\label{eq:L2BB}
\mathcal{L}^{(BB)}_{(2\text{-ii})}= \mathcal{F}_{\alpha\beta}
\varepsilon^{\mu_1\mu_2\mu_3\mu_4}\varepsilon^{\rho_1\rho_2}{}_{\mu_3\mu_4}
\partial_{\mu_1}B^{(\alpha)}_{\rho_1}\partial_{\mu_2}B^{(\beta)}_{\rho_2}\,,
\end{align}
which belongs in $\mathcal{L}^{(BB)}_{(2)}$ and has
\begin{align}\label{eq:L2BBex1}
\mathcal{F}_{\alpha\beta} = \delta_{\alpha\beta}B^{(\gamma)}\cdot B_{(\gamma)} +2 B_{(\alpha)}\cdot B_{(\beta)}.
\end{align}
An $\mathcal{L}^{(AB)}_{(2)}$ example with $M,N>1$ is given by
\begin{align}\label{eq:L2ABex1}
\mathcal{L}^{(AB)}_{(2\text{-i})} =\mathcal{F}_{\alpha\bar{\beta}}{}^{\mu_{1}\mu_{2}}\,
\varepsilon^{\rho_1\rho_{2}\nu_2\sigma}\varepsilon_{\sigma\mu_{1}\mu_{2}}{}^{\nu_{1}}\,
\partial_{\nu_{1}} B^{(\alpha)}_{\rho_{1}}A^{(\bar{\beta})}_{\nu_2\rho_2}\,,
\end{align}
where the expansion coefficients are set to
\begin{align}
\mathcal{F}_{\alpha\bar{\beta}}{}^{\mu_{1}\mu_{2}}=\mathcal{K}^{[\alpha_{1}\alpha_{2}]}_{\bar{\beta}}\,
B_{(\alpha_{1})}^{\mu_{1}}B_{(\alpha_{2})}^{\mu_{2}} f_{\alpha}\,.
\end{align}
Here, the $\mathcal{K}^{[\alpha_{1}\alpha_{2}]}_{\bar{\beta}}$  are arbitrary constants and
\begin{align}\label{eq:2ndexcoeff}
f_{\alpha}=B^{(\gamma)}\cdot B_{(\gamma)}+2 B_{(\alpha)}\cdot \sum_{\gamma=1}^{M}B_{(\gamma)}\,\quad \forall \alpha.
\end{align}
It is easy to check that the interactions (\ref{eq:L2BB}) and (\ref{eq:L2ABex1}) satisfy (\ref{eq:secondCER})
and thus do not propagate any ghost-like degrees of freedom.
We emphasize that this holds true because (\ref{eq:L2BBex1}) and (\ref{eq:2ndexcoeff})
have been tailored to be in agreement with the secondary constraint enforcing relation.

In closing this section we recall that, for interactions involving more than one Proca field,
one must always ensure that (\ref{eq:secondCER}) is satisfied.
This imposes highly non-trivial restrictions on the interactions.

\section{Relaxing the axioms: beyond Maxwell-Proca \label{sec:beyondMP}}
The completeness of the just proposed Maxwell-Proca theory rests on the four axioms
encoded in Definitions \ref{def:Maxwell}-\ref{def:MaxPro} of section \ref{sec:axiomas}:
\begin{enumerate}[i]
\item All vector fields are Abelian. \label{axiom1}
\item All vector fields are taken to be real. \label{axiom2}
\item The Lagrangian is at most first order. \label{axiom3}
\item We work on Minkowski spacetime. \label{axiom4}
\end{enumerate}
To conclude, we comment on the scenarios that unfold upon concrete relaxations of the assumed postulates.
In all cases, the materialization of our subsequent proposals into an exhaustive Lagrangian is non-trivial and lies beyond the scope of this paper.

{\it Generalization to non-Abelian fields.}
As is widely known, classical Yang-Mills theory~\cite{YM} can be regarded as an extension of Maxwell electrodynamics to non-Abelian vector fields.
Conversely, electromagnetism can be understood as a Yang-Mills theory with gauge group $U(1)$.
This relationship has already been exploited to construct generalized Yang-Mills theories
starting from specific non-linear electrodynamics~\cite{Goldin:2004ea}.
On the other hand, non-Abelian extensions of the Proca field are not possible.
Indeed and as already noted in~\cite{Heisenberg:2017mzp,Allys:2016kbq}, attempts at constructing a massive version of the Yang-Mills theory boil down to the consideration of multiple interacting Abelian Proca fields in the presence of a global rotational symmetry in the field space.
Putting together both observations, the non-Abelian analogue to the Maxwell-Proca theory could be dubbed Yang-Mills-Proca.
Notice that the constraint enforcing relations (\ref{eq:primCER}) and (\ref{eq:secondCER}), as well as the purely massive interactions $\mathcal{L}^{(BB)}$ in (\ref{eq:MPLintSchem}), are the same as in the Maxwell-Proca theory.

{\it Generalization to complex fields.}
While the extension of the massless sector to the complex field case is straightforward, the study of a complexified massive sector is not.
For example, allowing for Proca fields to be complex makes it possible to charge them under the fundamental representation
of the $U(1)$ gauge group of any of the Maxwell fields:
\begin{equation}
B_{\mu}\longrightarrow e^{ie\Phi}B_{\mu}, \qquad B_\mu^\ast\longrightarrow B_\mu^\ast e^{-ie\Phi},
\label{charged}
\end{equation}
with $e$ a coupling constant and $\Phi$ the same real scalar field that appeared in (\ref{eq:U1A}) before.
Then, one can generate new interaction terms by replacing partial derivatives $\partial_\mu$ acting on the Proca fields by covariant derivatives
$D_\mu:=\partial_\mu+ie A_\mu$, as long as the resulting term is invariant under the simultaneous transformations (\ref{eq:U1A}) and (\ref{charged}).
Observe that such interactions imply the Maxwell fields are no longer restricted to appear as part of a field strength.
This generalization does not affect the form of the constraint enforcing relations (\ref{eq:primCER}) and (\ref{eq:secondCER}).

{\it Generalization to higher order field theory.}
If the Maxwell-Proca theory is promoted to depend on arbitrarily high order time derivatives of the fields, i.e.
\begin{align}
\mathcal{L}_{\textrm{MP}}\longrightarrow\mathcal{L}_{\textrm{MP}}(A_\mu,B_\mu,\dot{A}_\mu,\dot{B}_\mu,\ddot{A}_\mu,\ddot{B}_\mu,\dots),
\label{highL}
\end{align}
the Lagrangian and Hamiltonian formalisms underlying Appendix \ref{app:GhostFreedom}
no longer constitute the suitable framework to study the constraint algebra.
Indeed, the development of an appropriate formalism has been investigated for a long time and continues to be pursued, see e.g.~\cite{HOFT}.
The foundations of a consistent approach amenable to geometrization were laid in~\cite{Gracia:1989ig},
based on the notions put forward in~\cite{BGPRR} for second order theories.
The main idea consists on introducing so-called intermediate spaces that connect the tangent bundles $\mathcal{T}^{2k-1}\mathcal{C}$,
natural to the Lagrangian analysis,
to the cotangent bundles $\mathcal{T}^\ast(\mathcal{T}^{k-1}\mathcal{C})$, which appear on the Hamiltonian side.
Here, $k=1,\dots,n$ for a $n$th order Lagrangian and $\mathcal{C}$ is the configuration space.
The intermediate spaces sustain a systematic inspection of the higher velocities and a step-by-step inference of the corresponding momenta.
Moreover, iterative algorithms that ensure the stability of the constraints in the theory are naturally defined in these spaces,
under mild requirements on the form of (\ref{highL}).
Obviously, the constraint enforcing relations (\ref{eq:primCER}) and (\ref{eq:secondCER}) do not apply to higher order cases.

{\it Generalization to curved backgrounds.}
Relaxing our last axiom \ref{axiom4} means allowing for general background configurations,
such as (anti-)de-Sitter or Friedmann-Robertson-Walker spacetimes.
The complete set of permitted metrics can be obtained by coupling the Maxwell-Proca theory to gravity.
As discussed elsewhere~\cite{ErrastiDiez:2019trb}, this leads to diverse physical applications.

Regarding free vector fields with a standard quadratic mass,
a consistent coupling to General Relativity is obtained by direct covariantization.
For instance, the Proca mass term in Minkowski space $m^2 \eta^{\mu\nu} B_{\mu}B_\nu$ is promoted to $m^2 g^{\mu\nu} B_{\mu}B_\nu$,
where $g^{\mu\nu}$ is (the inverse of) the dynamical spacetime metric.
For terms containing two or more derivatives, this procedure becomes ambiguous due to the presence of the derivatives $\partial_\mu$, which do not commute anymore when replaced by covariant derivatives $\nabla_\mu$.

When covariantizing the Maxwell-Proca action one needs to be particularly careful to not destroy the underlying constraint algebra.
This algebra should contain the constraints for the massless spin-two together with those of the Maxwell-Proca theory.
In other words, we expect the constraint enforcing relations (\ref{eq:primCER}) and (\ref{eq:secondCER})
to be supplemented by additional conditions coming from the gravitational sector.

In fact, already for the single Proca theory in~\cite{Tasinato:2014eka}, non-trivial curvature couplings appear in the action.
The precise form of these terms was motivated by focusing on the longitudinal mode of the vector field
and requiring that its couplings to gravity be the same as in Horndeski's theory, which is known to be ghost-free.
This is a necessary but not sufficient condition for a (first order) Lagrangian to propagate the correct number of degrees of freedom.

The authors of~\cite{Hull:2015uwa} took a more rigorous approach and initiated an ADM analysis~\cite{ADM} of the interactions proposed in~\cite{Tasinato:2014eka}.
Their results show that the non-minimal curvature couplings are precisely such that
both the $B_0$ component of the Proca field and the lapse function in the metric appear without time derivatives in the action.
This demonstrates that the Lagrangian in~\cite{Tasinato:2014eka} satisfies another necessary condition for consistency.
A full constraint analysis of the theory that provides a solid proof of ghost freedom, however, is still pending.
Based on these partial results for the single Proca field,
it is natural to expect that the ghost-free covariant version of our Maxwell-Proca theory will also contain non-minimal curvature couplings.

We finish by pointing out two recent developments pertinent to the beyond Maxwell-Proca scenarios above described.
First, it is interesting to note that a particular example of a consistent theory that simultaneously relaxes axioms \ref{axiom3} and \ref{axiom4}
arises in~\cite{deRham:2018svs}.
One of the decoupling limits of non-linear massive gravity~\cite{deRham:2010ik} there considered
distributes the theory's five propagating modes into two plus three degrees of freedom,
corresponding to a massless graviton and a Proca field, respectively.
The resulting vector field Lagrangian contains higher order terms and is defined on four-dimensional anti-de Sitter spacetime,
meanwhile it is ghost-free by construction, as proven in~\cite{Hassan:2011tf}.
Second, when relaxing axiom \ref{axiom4} and coupling the Maxwell-Proca theory to gravity, one should be careful not to miss
the kind of terms unveiled shortly after this work, in~\cite{GallegoCadavid:2019zke}. Such terms reduce to total derivatives in flat spacetime.

\section{Concluding remarks \label{sec:remarks}}

The key ideas behind the results here obtained are as follows.
We begin by formalizing the notion of a Proca field in Definition \ref{def:Proca},
putting it on the same footing as that of the renowned Maxwell field (see Definition \ref{def:Maxwell}).
Based on these two concepts, we construct the Maxwell-Proca Lagrangian:
the most general first-order theory in Minkowski spacetime for an arbitrary number of interacting Maxwell and Proca fields propagating the correct number of degrees of freedom.

In more detail, the Maxwell-Proca Lagrangian is most conveniently expressed as the sum of kinetic plus interaction terms (\ref{eq:MPLagSchem}).
The kinetic piece is (\ref{eq:MPLkinSchem}).
The interactions are obtained by systematically converting the uniquely defined primary constraint enforcing relations (\ref{eq:primCER})
---derived in the appendix (\ref{primcons})--- into as definite a Lagrangian as possible.
We distinguish two types of interactions in (\ref{eq:fulllag}): those that trivially satisfy (\ref{eq:primCER}) and those that don't.
The first set can succinctly be written as (\ref{eq:L0piece}), while the coefficients of the second set are made explicit in (\ref{eq:T1})-(\ref{eq:T4}).

Additionally, when multiple interacting Proca fields are considered,
this Lagrangian must be forced to satisfy the uniquely defined secondary constraint enforcing relations in (\ref{eq:secondCER})
---derived in the appendix (\ref{mainres})--- to avoid ghosts.
No further conditions on the Lagrangian apply because, once equation (\ref{eq:secondCER}) is fulfilled, the constraint algebra automatically closes at the tertiary level.
Notice that, in full generality, (\ref{eq:secondCER}) is a set of coupled non-linear partial differential equations whose solution is not known.
This explains our inability to turn (\ref{eq:secondCER}) into a Lagrangian building principle, unlike (\ref{eq:primCER}) before.

The vast and diverse applicability of the Maxwell-Proca theory is discussed and exemplified elsewhere~\cite{ErrastiDiez:2019trb}.

\begin{acknowledgments}
The authors thank Axel Kleinschmidt for a fruitful discussion, which gave us the idea for this work.
We are grateful to Chrysoula Markou for enlightening conversations in the early stages of the project.
VED thanks Nicol\'{a}s Coca L\'{o}pez for lucid explanations regarding superconductors.
JAMZ is thankful to  E. A. Ay\'{o}n-Beato and D. F. Higuita-Borja for the computing power granted at ZymboLab (CINVESTAV).
The work of JAMZ is partially supported by the ``Convocatoria Max-Planck-CONACyT 2017 para estancias postdoctorales'' fellowship.
This work is supported by a grant from the Max Planck Society.\\
VED lovingly dedicates this work to her mother.
\end{acknowledgments}

\appendix

\begin{widetext}
\begin{table}[!h]
\centering
\begin{tabular}{|l|l|l|l|l|}
\cline{2-5}
\multicolumn{1}{c|}{} & \multicolumn{3}{c|}{\bf{On-shell}} & \multirow{2}*{\hspace{0.5cm}\bf{Off-shell}} \\ \cline{2-4}
\multicolumn{1}{c|}{} & \,\,\emph{Primary} & \,\,\emph{Secondary} & \,\,\emph{Tertiary}  & \\ \hline
Euler-Lagrange eqs.  & \,\,$M_0-M_1=3(N+M)$\hspace{0.4cm}(\ref{primEL})\,\,   & \,\,$K_1-M_2\overset{!}{=}0$ \hfill
& \,\,$K_2-M_3=M$\hspace{0.2cm} (\ref{terteom})\,\, & \,\,$\emptyset$  \\ \hline
Identities & \,\,$\emptyset$\hfill    & \,\,$L_2=0$\hfill
& \,\,$L_3=0$\hfill & \specialcell{  \,\,$L_1=N$ \hspace{0.4cm}(\ref{Bianchiid})\,\,\\
\,\,$g=N$\hfill (\ref{gaugeid})\,\, } \\ \hline
Lagrangian constraints & \,\,$K_1=M$\hfill (\ref{primlag})\,\,  & \,\,$K_2=M$\hspace{1.0cm}(\ref{seclag})\,\,
& \,\,$K_3=0$\hfill  & \,\,$\emptyset $  \\ \hline
\end{tabular}
\captionsetup{justification=centerlast,width=2\columnwidth}
\caption{\label{table1}Summary of the Lagrangian composition of the Maxwell-Proca theory with $N$ Maxwell and $M$ Proca fields.
The table shows the number $g$ of gauge identities present,
together with the independent Euler-Lagrange equations of motion, $L$ Bianchi identities and $K$ Lagrangian constraints.
When possible, these are further classified into primary, secondary and tertiary types.
The explicit form of all the tabulated quantities can be found in the quoted formulae.}
\end{table}
\end{widetext}

\section{Degree of freedom count \label{app:GhostFreedom}}
In this appendix, we provide the explicit count of the degrees of freedom propagated by the Maxwell-Proca theory.
We thus demonstrate the absence of ghosts claimed around equation (\ref{eq:MPLintSchem}) in Sec.~\ref{sec:axiomas}.
Namely, given the Lagrangian constructed in Sec.~\ref{sec:construction}, this count ensures that Definitions \ref{def:Maxwell}-\ref{def:MaxPro} in Sec.~\ref{sec:axiomas} hold true.
Indeed, in sections \ref{app:LagCons} and \ref{app:HamCons} we obtain the results summarized in tables \ref{table1} and \ref{table2}, respectively.

Our starting point is the Maxwell-Proca action constructed in Sec.~\ref{sec:construction}, equations (\ref{eq:fulllag})-(\ref{eq:T4}).
For our present purposes, it suffices to consider this action in the schematic form
\begin{align}
S=\int_{\mathbb{R}\times \mathbb{R}^3}\textrm{d}^{3+1}x\,\, \mathcal{L}, \qquad \mathcal{L}=\mathcal{L}(Q^I,\dot{Q}^I). \label{maxpract}
\end{align}
Here, the $\{Q\}$'s are the $M_0=4(N+M)$ field variables that label the points in the configuration space $\mathcal{C}$ of the system,
\begin{align}
Q^I&=(A^{(1)\mu},\dots,A^{(N)\mu},B^{(1)\mu},\dots,B^{(M)\mu}) \label{gencoord}
\end{align}
with $I=1,\dots, M_0$, and the dot denotes derivation with respect to time $\dot{Q}^I:=\partial_0 Q^I$.
The velocities $\{\dot{Q}\}$ span the tangent space $\mathcal{TC}$ of the configuration space $\mathcal{C}$.
In all generality, we take $\mathcal{C}$ to be a differentiable Banach manifold.
The Lagrangian $\mathcal{L}$ depends on the spatial derivatives of the field variables as well, $\mathcal{L}=\mathcal{L}(\partial_i Q^I)$.
However, such dependencies are not relevant to the upcoming degree of freedom count and hence we omit them.
Note that our $\mathcal{L}$ does not depend on time explicitly.
This simplifies a bit the subsequent algebra, but it doesn't introduce nor remove any conceptual feature compared to the generic scenario
$\mathcal{L}=\mathcal{L}(t,Q^I,\dot{Q}^I)$.

Our end point will be the well-known classical Hamiltonian formula that enumerates propagating degrees of freedom:
\begin{align}
n_{\textrm{dof}}=M_0-N_1-\frac{N_2}{2}, \label{ndof}
\end{align}
where $N_1$ and $N_2$ stand for the number of first and second class constraints, respectively.
Recall that, for physically sensible theories, $N_2$ is always an even integer.
We shall find $N_1=2N$ by construction and restrict the form of the Lagrangian through the demand $N_2\overset{!}{=}2M$,
in order to obtain the desired count of degrees of freedom: two per Maxwell field and three per Proca field.

Needless to say, the formalism underlying this appendix is not new.
It has been long established that any field theory, such as the Maxwell-Proca theory here studied,
is always described by a singular Lagrangian ---for example, see~\cite{Henneaux:1992ig}.
The equivalence between the Lagrangian and Hamiltonian pictures~\cite{Pons} implies that any field theory is always a constrained Hamiltonian system.
Then, the famous Dirac-Bergmann (Hamiltonian) theory of constraints~\cite{DiracBergmann,AndBergmann} ---which was geometrized in~\cite{Gotay}---,
provides the canonical formalism to study such theories.
In particular, it gives rise to our end point formula (\ref{ndof}).

Our appendix describes the fastest and easiest way to obtain the results of interest.
We predominantly make use of the Lagrangian formalism for two fundamental reasons.
First and foremost, this allows for the construction of the Maxwell-Proca theory.
We anticipate that the uniquely defined primary constraint enforcing relations in (\ref{primcons}) are rewritten as equation (\ref{eq:primCER}),
which is then systematically converted into as definite a Lagrangian as possible in section \ref{sec:PrimLevel}.
Then, the also uniquely defined secondary constraint enforcing relations in (\ref{mainres}) are rewritten as equation (\ref{eq:secondCER}).
This latter set of equations is left as an additional requirement on the Lagrangian that must be satisfied to ensure ghost-freedom.
Second, the applications of the Maxwell-Proca theory we propose in~\cite{ErrastiDiez:2019trb},
as well as the cosmological applications that motivated the very notion of a generalized Proca field~\cite{Tasinato:2014eka}
use the Lagrangian (and not the Hamiltonian) of the theory as their starting point.
We choose to make use of the Hamiltonian formalism only to explicitly count the degrees of freedom propagated by the Maxwell-Proca theory,
thereby showing it is free of ghosts.
The Hamiltonian itself is not worked out because this is not required for the counting process.

The fact that the Maxwell-Proca theory is a gauge theory introduces subtleties on both the Lagrangian and Hamiltonian side of the story.
We will discuss and reference this important point where appropriate.
For the benefit of the reader, the appendix is written as an up-to-date and self-contained partial review of the theory of singular Lagrangians and constrained Hamiltonians in the presence of gauge symmetries.

\subsection{Lagrangian constraint analysis \label{app:LagCons}}
As already anticipated, we begin the study of the constraint structure of the Maxwell-Proca theory in the Lagrangian formalism.
In this section, we obtain the results conveyed in table \ref{table1}.
Namely, we count the number of independent Euler-Lagrange equations of motion, Bianchi identities and Lagrangian constraints present in the theory
and classify them into primary, secondary and tertiary.
We show that no tertiary Lagrangian constraints appear and hence no quaternary level arises.
We also find the gauge identities in the theory.

What follows is not just a counting exercise, though.
In the process, we will find the necessary and sufficient conditions to make our count match that of the auxiliary theory
\begin{align}
\mathcal{L}_{\textrm{aux}}:=-\frac{1}{4}A_{\mu\nu}^{(\overline{\alpha})}A^{\mu\nu}_{(\overline{\alpha})}
-\frac{1}{4}B_{\mu\nu}^{({\alpha})}B^{\mu\nu}_{({\alpha})}-\frac{m^2}{2}B_{\mu}^{(\alpha)}B^{\mu}_{(\alpha)}. \label{mp}
\end{align}
This is the theory of $N$ Maxwell and $M$ Proca {\it free} fields, where for simplicity we have taken the mass $m\neq0$ of all the Proca fields to be the same.
The aim is to show that the (self-)interactions between all $N+M$ fields that we proposed  in (\ref{eq:fulllag}) do not introduce further unphysical
degrees of freedom, compared to $\mathcal{L}_{\textrm{aux}}$.
We will see this does not happen automatically: for $M>1$, the constraint enforcing relation (\ref{eq:secondCER0}) must be fulfilled.

Because we adopt a coordinate-dependent approach, the subsequent analysis depends on the form of the Lagrangian:
the number of identities and Lagrangian constraints varies with the field basis
(unlike the number of degrees of freedom, which is basis-independent).
The results here reported correspond to the mass eigenbasis formulation of the Maxwell-Proca theory,
which was introduced in Sec \ref{sec:MassEigenbasis}.

Our analysis is largely based on the iterative algorithm presented in~\cite{Kamimura:1981fe}.
However, there exists a vast literature on the topic that could be consulted equivalently:~\cite{Lagcons}, to mention but a few of the standard references.

{\it Iterative analysis.}
Given the classical field theory Lagrangian density (\ref{maxpract}) of interest, the principle of stationary action leads to the following
Euler-Lagrange equations of motion:
\begin{align}
E_J:= \ddot{Q}^IW_{IJ}+\alpha_J=0 \qquad \forall J, \label{ELeq}
\end{align}
where we have defined
\begin{align}
W_{IJ}:=\frac{\partial^2 \mathcal{L}}{\partial \dot{Q}^I\partial \dot{Q}^J}, \qquad
\alpha_J:=\dot{Q}^I\frac{\partial^2\mathcal{L}}{\partial Q^I\partial \dot{Q}^J}-\frac{\partial\mathcal{L}}{\partial Q^J}.\label{primhess}
\end{align}
In (\ref{ELeq}), the right-hand side is an on-shell statement, while $W_{IJ}$ is a $M_0\times M_0$ matrix known as the Hessian.
By construction\footnote{At this point,
we explicitly verify that the Maxwell-Proca interaction terms in equation (\ref{eq:fulllag}) satisfy the requirement
of ghost-freedom at the primary level.
Recall that these interactions were generated  in Sec.~\ref{sec:PrimLevel}, using a systematic procedure based on pair-wise antisymmetrization
of the derivatives $\partial_\mu Q^I\subset\mathcal{L}$ using Levi-Civita symbols.}, our $\mathcal{L}$ is such that
\begin{align}
W_{IJ}&\equiv 0  \label{primcons}
\end{align}
for all $J$ and for $I=4z-3$, with $z=1,\dots,N+M$.
The above is the necessary and sufficient condition to ensure ghost-freedom of the theory at the primary level.
In the language of the main text, this condition reads  (\ref{eq:primCER}).
It was until now believed~\cite{Heisenberg:2017mzp,Allys:2016kbq} that (\ref{primcons}) is the only condition needed to avoid ghosts.
We will show that this expectation is too naive and that, for $M>1$, (\ref{primcons}) must be supplemented by an additional condition (arising at the secondary level) in order to truly remove ghosts.

Note that (\ref{primcons}) makes the rank of the Hessian be smaller than its dimension and so its determinant vanishes.
This implies that the Lagrangian density under consideration is singular, as we said it should for a field theory.
What is more, (\ref{primcons}) yields a Hessian of rank $3(N+M)$.
This is what is needed to reproduce the primary level of (\ref{mp}).
The desired rank reduction of the Hessian can only happen by setting entire columns (or, equivalently, rows)
of this matrix to zero for any first order Lagrangian, i.e.~as in (\ref{primcons}).
Rank reductions in the form of $N+M$ columns being linear combinations of the remaining set of linearly independent $3(N+M)$ columns
can be absorbed into a field redefinition.

Given (\ref{primcons}), it is clear that not all the equations in (\ref{ELeq}) are independent second order differential equations;
only $M_0-M_1$ number of them are.
The remaining $M_1$ equations do not involve the accelerations $\{\ddot{Q}\}$.
Instead, they are relations between the generalized field variables $\{Q\}$ and their time derivatives $\{\dot{Q}\}$.
Here,
\begin{align}
M_1&:= \textrm{dim}(W_{IJ})-\textrm{Rank}(W_{IJ})=N+M.
\end{align}
Among these $M_1$ relations, we will further distinguish between Bianchi identities and primary Lagrangian constraints,
whose number we denote by $L_1$ and $K_1$, respectively.

In order to split (\ref{ELeq}) into the above mentioned three subsets of equations,
one must first obtain the $M_1$ linearly independent zero vectors $\{\gamma^I\}$ associated to the Hessian:
\begin{align}
 \gamma_a{}^IW_{IJ}=0, \qquad a=1,2,\dots,M_1. \label{gammprim}
\end{align}
These null vectors must be chosen so that they satisfy
\begin{align}
\gamma_a{}^I\gamma_I{}^b=\delta_a{}^b. \label{orth}
\end{align}
Additionally, one must also compute an auxiliary $M_0\times M_0$ matrix $M^{IJ}$.
This matrix always exists and is uniquely determined from the relations:
\begin{align}
W_{IJ}M^{JK}-\delta_I{}^K+\gamma_I{}^a\gamma_a{}^K=0, \qquad M^{IJ}\gamma_J{}^a=0.\label{MIJ}
\end{align}
Then, the $M_0-M_1$ independent second order (primary) Euler-Lagrange equations are the non-zero subset in
\begin{align}
(\ddot{Q}^J+\alpha_IM^{IJ})W_{JK}=0, \label{primEL}
\end{align}
while the $M_1$ relations between the $Q$'s and their velocities are
\begin{align}
\varphi_a:=\gamma_a{}^I\alpha_I=0 \qquad \forall a. \label{M1rel}
\end{align}
Notice that the right-hand side is, in principle, an on-shell equality.
For the Maxwell-Proca theory, $N$ number of the $\varphi$'s are identically zero:
\begin{align}
\varphi_{\overline{\alpha}}\equiv0 \qquad \forall \overline{\alpha}=1,\dots,N.\label{Bianchiid}
\end{align}
These are the $L_1=N$ Bianchi identities associated to the Maxwell fields, which will be later related to first class constraints.
It is important to realize that (\ref{Bianchiid}) holds true off-shell: it does not require the right-hand side of (\ref{M1rel}).
In general, any identity at the primary level is off-shell by definition.
On the other hand, the imposition that the remaining $\varphi$'s vanish,
\begin{align}
\varphi_{\alpha}\approx 0 \qquad \forall \alpha=1,\dots,M,\label{primlag}
\end{align}
is an on-shell statement that
gives rise to $K_1=M$ primary Lagrangian constraints on the Proca fields\footnote{Notice that in
(\ref{Bianchiid})-(\ref{primlag}) we reuse
the field labels ($\overline{\alpha}$, $\alpha$) that in the main text enumerate Maxwell and Proca fields, respectively.
Throughout the Lagrangian analysis, these labels enumerate Bianchi identities and primary Lagrangian constraints instead.
Since these are intrinsically associated to the Maxwell and Proca fields, the slight abuse in the notation serves as a reminder of their origin.
\label{footalph}}.
These will be later connected with second class constraints.
It is worth pointing out that, in agreement with the standard notation, ``$\approx$'' stands for a weak equality.
In more detail, the requirement that those $\varphi$'s which are not identities vanish defines a subspace $\mathcal{C}_0\subseteq \mathcal{C}$
in the configuration space, generally referred to as the constraint surface.
A weak equality is one which only holds true in $\mathcal{C}_0$, not necessarily in the entire of $\mathcal{C}$.
We here finish the study of the primary level of the Maxwell-Proca theory.

A significant remark on nomenclature follows.
Equation (\ref{Bianchiid}) is referred to as Bianchi identities because these are the natural generalization of
the familiar Bianchi identity in classical electromagnetism to the full Maxwell-Proca theory.
In the language of differential forms, one can succinctly write these full Maxwell-Proca Bianchi identities as $dF-\mathcal{J}=0$,
where $d$ denotes exterior derivative, $F$ is the field strength of the usual Maxwell field
and $\mathcal{J}$ is the most general possible source term within the defining axioms of the theory.
That is, $\mathcal{J}$ consists of a set of sources coming from the non-linearities of the usual Maxwell field, from additional Maxwell fields and from all the Proca fields.
In the well-known Maxwell electrodynamics limit ($N=1$ and $M=0$ with no interaction terms),
$\mathcal{J}=0$ and the usual Bianchi identity $dF=0$ is recovered.

Before proceeding to the secondary level, we will briefly discuss the increased algorithm needed for reducible constrained systems.
These are theories for which the on-shell relations (\ref{M1rel}) are not functionally independent.
In our setup, we encounter this feature in two instances:
\begin{enumerate}[i]
\item If working with a mixed eigenstate basis,
obtained from the mass eigenbasis $\{A,B\}$ by simple linear field redefinitions. \label{casei}
\item If studying the charged complex Proca proposal outlined in Sec \ref{sec:beyondMP}. \label{caseii}
\end{enumerate}
For any reducible theory, (\ref{M1rel}) satisfy $L_1$ relations of the form
\begin{align}
Z_{\overline{\alpha}}{}^{a}\varphi_{a}\equiv0 \qquad \textrm{with }Z_{\overline{\alpha}}\in \mathcal{TC} \quad \forall {\overline{\alpha}}.
\label{redeq}
\end{align}
The above are the analogues of the Bianchi identities (\ref{Bianchiid}) of the irreducible case.
The case \ref{casei} is trivial, in the sense that (\ref{redeq}) amount to linear relations between the reducible constraints $\{\varphi\}$.
On the other hand, for the case \ref{caseii}, the $Z$'s include differential operators and can be determined iteratively order by order\footnote{Although we are not concerned with such involved scenarios, in~\cite{Diaz:2017tmy} it is noted that there exist theories where infinitely many iterations would be required to obtain the $Z$'s.}, as explained in~\cite{Diaz:2017tmy}.
Once that (\ref{redeq}) has been established, only the $M_1-L_1$ irreducible Lagrangian constraints $\{\varphi\}$ analogue to (\ref{primlag})
are to be used to continue with the iterative algorithm.

Going back to the Maxwell-Proca theory,
the demand that (\ref{primlag}) holds true at all times (i.e.~$\dot{\varphi}_\alpha\approx0$ for all $\alpha$) initiates the study of the secondary level\footnote{For clarity, we will use a notation where
tilde quantities correspond to the secondary level (or second iteration in the constraint analysis algorithm in~\cite{Kamimura:1981fe})
and hat quantities belong to the tertiary level (third iteration).}, which is an innately on-shell level.
Specifically, it gives rise to $K_1=M$ secondary equations of motion:
\begin{align}
\widetilde{E}_\beta:=\ddot{Q}^I\gamma_I{}^\alpha \widetilde{W}_{\alpha\beta}+\widetilde{\alpha}_\beta=0 \qquad \forall \beta, \label{secEoM}
\end{align}
where we have defined
\begin{align}
\widetilde{W}_{\alpha\beta}:=\gamma_\alpha{}^I\frac{\partial \varphi_\beta}{\partial\dot{Q}^I}, \quad
\widetilde{\alpha}_\alpha:=-\alpha_I M^{IJ}\frac{\partial\varphi_\alpha}{\partial \dot{Q}^J}+\dot{Q}^I\frac{\partial\varphi_\alpha}{\partial Q^I},
\label{secalhes}
\end{align}
in close analogy to the primary quantities in (\ref{primhess}).
As happened with (\ref{ELeq}) before, generally not all the above secondary equations of motion are independent second order differential equations.
Instead, they can again be divided into three subsets: $K_1-M_2$ secondary Euler-Lagrange equations,
$L_2$ identities and $K_2$ secondary Lagrangian constraints, with
\begin{align}
M_2:=\textrm{dim}(\widetilde{W}_{\alpha\beta})-\textrm{Rank}(\widetilde{W}_{\alpha\beta}).
\end{align}

In order for the Maxwell-Proca theory to have the desired Lagrangian conformation (i.e.~that of the auxiliary theory in (\ref{mp})),
$M_2=K_1$ is needed, implying that there are no independent secondary Euler-Lagrange equations.
It is easy to see that this is uniquely achieved by setting
\begin{align}
\widetilde{W}_{\alpha\beta}\overset{!}{\equiv}0 \qquad \forall\alpha,\beta, \label{mainres}
\end{align}
which is the sufficient and necessary condition that ensures ghost-freedom at the secondary level.
Fulfillment of (\ref{mainres}) implies the constraint enforcing relations (\ref{eq:secondCER}) on the Maxwell-Proca interaction terms.
Written in this manner, it is easy to see that
the above identical vanishing happens by construction for any number of Maxwell fields and when a single Proca field is considered.
In other words, (\ref{mainres}) trivially holds true if $M\leq1$.
However, when two or more Proca fields are present in the theory, (\ref{mainres}) ---equivalently, (\ref{eq:secondCER})--- non-trivially restricts the proposed interaction terms.
We highlight that this is precisely the anticipated condition that was missed out in all~\cite{Heisenberg:2017mzp,Allys:2016kbq} and that supplements the known condition (\ref{primcons}).
To avoid the propagation of ghosts in any multi-Proca setup, both (\ref{primcons}) and (\ref{mainres}) must hold true.
As noted in the main text, we regard this realization as our main result.

Given (\ref{mainres}), repetition of the logic exposed for the primary level requires us to obtain
the $M_2$ null vectors $\{\widetilde{\gamma}^\alpha\}$\footnote{Note that the $\widetilde{\gamma}$'s are
defined in $\mathcal{C}_0$. In particular, they are only determined modulo the $\varphi$'s.} associated to $\widetilde{W}_{\alpha\beta}$,
\begin{align}
\widetilde{\gamma}_r{}^\alpha\widetilde{W}_{\alpha\beta}\approx 0, \qquad r=1,2,\dots,M_2,
\end{align}
in such a way that they satisfy the orthonormality relation
$\widetilde{\gamma}_r{}^\alpha\widetilde{\gamma}_\alpha{}^s=\delta_r{}^s$.
Further, it is easy to see that the always extant $M_1\times M_1$ auxiliary matrix $\widetilde{M}^{\alpha\beta}$, which is analogue to (\ref{MIJ}) at the
secondary level and which is uniquely determined from
\begin{align}
\widetilde{W}_{\alpha\beta}\widetilde{M}^{\beta\gamma}-\delta_\alpha{}^\gamma+\widetilde{\gamma}_\alpha{}^r\widetilde{\gamma}_r{}^{\gamma}=0,
\qquad \widetilde{M}^{\alpha\beta}\widetilde{\gamma}_\beta{}^r=0,
\end{align}
identically vanishes in this case: $\widetilde{M}^{\alpha\beta}\equiv0$.
Subsequently, the $M_2$ secondary relations defined in $\mathcal{C}_0$ among the generalized coordinates and their time derivatives are given by\footnote{For reducible theories, the increased iterative algorithm described around (\ref{redeq}) should be applied at the secondary level as well.}
\begin{align}
\phi_r:=\widetilde{\gamma}_r{}^\alpha\widetilde{\alpha}_\alpha\approx0 \qquad \forall r. \label{phir}
\end{align}
We see that none of the $\phi$'s vanish identically, implying $L_2=0$.
To obtain the $K_2=M_2$ secondary Lagrangian constraints in the theory, one must indeed demand that they are zero:
\begin{align}
\phi_{r}\overset{1}{\approx} 0 \qquad \forall r. \label{seclag}
\end{align}
We will soon see that these relate to second class constraints for the Proca fields.
For clarity, we have here introduced the symbol ``$\overset{1}{\approx}$'' to indicate that
the requirement that the $\phi$'s vanish defines a subspace $\mathcal{C}_1\subseteq \mathcal{C}_0$ in the constraint surface.
As a result, a  weak equality that only holds true in $\mathcal{C}_1$ (and not necessarily in $\mathcal{C}_0$ or the entire of $\mathcal{C}$)
shall be denoted by $\overset{1}{\approx}$ in the following.

One more iteration is required.
Again, all statements at this tertiary level hold on-shell only.
Demanding that (\ref{seclag}) holds true at all times (namely $\dot{\phi}_r\approx0$ for all $r$), $M_2=M$ number of tertiary equations of motion are obtained:
\begin{align}
\widehat{E}_s:=\ddot{Q}^I\gamma_I{}^\alpha\widetilde{\gamma}_\alpha{}^r\widehat{W}_{rs}+\widehat{\alpha}_s=0, \label{terteom}
\end{align}
with $\widehat{W}_{rs}$ and $\widehat{\alpha}_r$, the tertiary counterparts of (\ref{secalhes}) before,
defined as
\begin{align}
\widehat{W}_{rs}&:=\widetilde{\gamma}_r{}^\alpha\gamma_\alpha{}^I\frac{\partial \phi_s}{\partial \dot{Q}^I}
\end{align}
and
\begin{align}
\widehat{\alpha}_r&:=-\widetilde{\alpha}_\alpha\widetilde{M}^{\alpha\beta}\gamma_\beta{}^I\frac{\partial \phi_r}{\partial\dot{Q}^I}
-\alpha_IM^{IJ}\frac{\partial \phi_r}{\partial\dot{Q}^J}+\dot{Q}^I\frac{\partial\phi_r}{\partial Q^I},
\end{align}
respectively.
The $M_2$ equations (\ref{terteom}) can generically be divided into
$M_2-M_3$ independent tertiary Euler-Lagrange equations, $L_3(\leq M_3)$ identities and $M_3-L_3$ tertiary Lagrangian constraints,
where
\begin{align}
M_3:=\textrm{dim}(\widehat{W}_{rs})-\textrm{Rank}(\widehat{W}_{rs}).
\end{align}
However, $M_3=0$ for the Maxwell-Proca theory.
This implies that all equations in (\ref{terteom}) are independent tertiary Euler-Lagrange equations.
It also means $L_3=0=K_3$, which signals the termination of the iterative procedure:
an on-shell quaternary level would unfold from guaranteeing the stability of the $K_3$ tertiary Lagrangian constraints, but there are none.

{\it Gauge identity analysis.}
We turn our attention to gauge identities, which are known to escape the above algorithm and require separate consideration.
We shall use $g$ to denote their number.
The key point is to notice that the subset of $N$ number of $\alpha$'s defined in (\ref{primhess})
stemming from the Maxwell fields vanish identically:
\begin{align}
\alpha_{\overline{\alpha}}\equiv0 \qquad \forall\overline{\alpha}. \label{gaugeid}
\end{align}
This is obvious from their very definition, once one takes into account that our theory depends on the Maxwell fields only through their field strengths,
as pointed out  in Sec.~\ref{sec:MassEigenbasis}.
This is clearly an off-shell identity, since it is independent of (\ref{ELeq}).
Matter of fact,  (\ref{gaugeid}) are precisely the $g=N$ gauge identities we were looking for.
We will soon link them with first class constraints.
Notice that the Bianchies in (\ref{Bianchiid}) are projections of the gauges in (\ref{gaugeid})
to the kernel of the Hessian $W_{IJ}$ defined in (\ref{primhess}).
In this way, the Bianchi identities can be viewed as relics of the gauge invariance of the theory.
Accordingly, we will shortly relate them to first class constraints as well\footnote{At this point,
the reader may benefit from a lightning review of the familiar case of Maxwell electrodynamics.
Calling $(\vec{E},\vec{B})$ the electric and magnetic fields,
the Bianchi identities read $\vec{\nabla}\times \vec{E}+\dot{B}=0$ and $\vec{\nabla}\cdot\vec{B}=0$.
They are identically solved by $\vec{E}=-\vec{\nabla}\phi-\dot{\vec{A}}$ and $\vec{B}=\vec{\nabla}\times\vec{A}$, for any $(\phi,\vec{A})$.
The gauge symmetry tells us that not only these off-shell identities,
but also the Maxwell Lagrangian itself and the on-shell electric and magnetic Gauss' laws following from it
are invariant under $\vec{A}\rightarrow\vec{A}+\vec{\nabla}\lambda$ and $\phi\rightarrow\phi-\dot{\lambda}$,
with $\lambda$ an arbitrary smooth function of spacetime.}.

We have now fulfilled our initial promise and completed the derivation of table \ref{table1}.

\subsection{Hamiltonian constraint analysis \label{app:HamCons}}

Next, we focus on the study of the Hamiltonian constraints in the Maxwell-Proca theory.
In this section, we derive the results summarized in table \ref{table2}.
In more detail, we first relate the identities and Lagrangian constraints we found in the previous section to the primary and secondary Hamiltonian constraints in the Dirac-Bergmann formalism~\cite{DiracBergmann,AndBergmann}.
We loosely refer to these as Dirac primaries and secondaries, respectively.
Our analysis is primarily based on~\cite{Kamimura:1981fe} again.
Then, we reclassify these Dirac constraints as first or second class.
To this aim, we stick to~\cite{Date:2010xr}.
We restate, however, that an extensive bibliography exists elaborating on this topic ---for example, see~\cite{hamclass} and references therein.
In this way, we obtain the number of propagating degrees of freedom in the theory, according to (\ref{ndof}).
\begin{table}[!h]
\centering
\begin{tabular}{l|c|c|}
\cline{2-3} & \textsc{First class  } & \textsc{Second class  }  \\
\cline{1-3}
\multicolumn{1}{|l|}{Primaries \hfill(\ref{dirprim})\,\,} & \,\,$N$\hspace{0.4cm} $\widehat{\Psi}$ in (\ref{prim12})\,\,
& \,\,$M$\hspace{0.4cm} $\widetilde{\Psi}$ in (\ref{prim12})\,\, \\ \cline{1-3}
\multicolumn{1}{|l|}{Secondaries \,\,\hspace{0.0cm} (\ref{dirsec})\,\,} & \,\,$N$\hspace{0.4cm} $\widehat{\chi}$ in (\ref{sec12})\,\,
& \,\,$M$\hspace{0.4cm} $\widetilde{\chi}$ in (\ref{sec12})\,\, \\ \cline{1-3}
& $N_1=2N$ & $N_2=2M$  \\ \cline{2-3}
\end{tabular}
\captionsetup{justification=centerlast}
\caption{\label{table2}Number of Hamiltonian constraints present in the Maxwell-Proca theory with $N$ Maxwell and $M$ Proca fields, according to their type.
Their explicit form can be found in the quoted formulae. The non-singular part
of the constraint algebra is given by the Poisson bracket matrix (\ref{overlineDelta}).}
\end{table}

As already mentioned, the Hamiltonian formalism is an equivalent reformulation of the Lagrangian formalism in section \ref{app:LagCons}.
Its central object of study is the Hamiltonian density $\mathcal{H}$, whose explicit form we do not work out here
(as it is not required to count the propagating degrees of freedom).
The interested reader can consult in e.g.~\cite{SudMuk} the procedure to determine the $\mathcal{H}$ following from (\ref{maxpract}).
Formally, $\mathcal{H}$ relates to the Lagrangian density $\mathcal{L}$ as
\begin{align}
\mathcal{H}:=\dot{Q}^IP_I-\mathcal{L}(Q^I,\dot{Q}^I).
\end{align}
Generically, $\mathcal{H}$ is a function of three types of variables: the generalized coordinates $Q^I$ introduced in (\ref{gencoord}),
their time derivatives $\dot{Q}^I$ and the momenta $P_I$.
The $\{P\}$'s span the cotangent space $\mathcal{T}^\ast\mathcal{C}$ of the configuration space $\mathcal{C}$.
For clarity, we note that $P_I$ encodes both the momenta $\mathcal{P}_\mu^{(\overline{\alpha})}$
associated to the Maxwell fields and the momenta $\Pi_\mu^{(\alpha)}$ of the Proca fields:
\begin{align}
P_I&=(\mathcal{P}_\mu^{(1)},\dots,\mathcal{P}_\mu^{(N)},\Pi_\mu^{(1)},\dots,\Pi_\mu^{(M)}).
\end{align}
We denote as $\mathbf{R}$ the $(3M_0)$-dimensional space comprising $\mathcal{C}$, $\mathcal{TC}$ and $\mathcal{T}^\ast\mathcal{C}$.

The Hamiltonian density is independent of the $\dot{Q}$'s when restricted to the subspace $\mathbf{R}^0\subseteq \mathbf{R}$
defined by\footnote{The reader should not confuse the weak equality of the Lagrangian formalism,
defining $\mathcal{C}_0\subseteq\mathcal{C}$, with the weak equality of the Hamiltonian formalism, which defines $\mathbf{R}^0\subseteq \mathbf{R}$.}
\begin{align}
P_I-\frac{\partial \mathcal{L}}{\partial \dot{Q}^I}\approx 0.
\end{align}
In the case of the Maxwell-Proca theory of our interest, only $M_1=N+M$ of the above relations are independent.
These are the Dirac primary constraints $\Psi_a$.
We shall only be concerned with their determination modulo the soon to be introduced Poisson bracket operation, so we define them as
\begin{align}
\Psi_a:=\gamma_a{}^I\left(P_I-\frac{\partial \mathcal{L}}{\partial \dot{Q}^I}\right). \label{dirprim}
\end{align}
The Dirac primaries relate to the Lagrangian quantities $\{\varphi\}$ defined in (\ref{M1rel}) as
\begin{align}
\dot{\Psi}_a\approx \varphi_a, \label{primprim}
\end{align}
where the weak equality means this holds true in $\mathbf{R}^0$ only.
The demand that the Dirac primaries be stable,
\begin{align}
\dot{\Psi}_a \overset{1}{\approx} 0 \qquad \forall a, \label{dpsi0}
\end{align}
connects them with all the Bianchi identities in (\ref{Bianchiid}), the primary Lagrangian constraints in (\ref{primlag}) and the gauge identities in (\ref{gaugeid}).
On the Hamiltonian side, (\ref{dpsi0}) defines the subspace $\mathbf{R}^1\subseteq \mathbf{R}^0$.

The Dirac secondary constraints $\chi_a$ are defined as the $M_1=N+M$ independent relations between the $Q$'s and the $P$'s that vanish in $\mathbf{R}^1$ and that are independent of the velocities $\{\dot{Q}\}$ in $\mathbf{R}^0$.
As long as the Poisson bracket is concerned, it suffices for us to choose them in the rather obvious way
\begin{align}
\chi_a:=\dot{\Psi}_a. \label{dirsec}
\end{align}
A subset of $M$ Dirac secondaries relates to the Lagrangian objects $\{\phi\}$ defined in (\ref{phir}) as
\begin{align}
\dot{\chi}_r\approx \phi_r. \label{secsec}
\end{align}
The time derivative of the remaining $N$ Dirac secondaries is weakly vanishing, which can be viewed as the Hamiltonian reflection of the
intrinsic stability of the $L_1=N$ Bianchi identities in (\ref{Bianchiid}).
The requirement that the $\chi$'s be stable gives rise to the Lagrangian secondaries in (\ref{seclag}).

Having found the Dirac primary and secondary constraints in the theory, we proceed to their splitting into first and second class.
As a reminder, first class constraints are those which have a weakly vanishing Poisson bracket with all Dirac constraints.
The constraints not having this property are second class.
Equivalently, one can think of the first (second) class constraints as those which
do (don't) form a closed constraint algebra in Poisson brackets.
Recall as well that, in canonical (or Darboux) coordinates\footnote{By definition, these are natural orthonormal coordinates on $\mathbf{R}^0$.} $(Q^I,P_I)$, the Poisson bracket of two functions $f$ and $g$ in $\mathbf{R}^0$ takes the form
\begin{align}
\{f,g\}\approx \frac{\partial f}{\partial Q^I}\frac{\partial g}{\partial P_I}-\frac{\partial f}{\partial P_I }\frac{\partial g}{\partial Q^I},
\qquad \forall f,g\in\mathbf{R}^0.
\end{align}

In order to divide the Dirac primaries into first and second class,
we need to consider the matrix of the Poisson brackets of the constraints in the theory:
\begin{align}
\Delta:=\left(
\begin{array}{cccc}
\{\Psi_1,\Psi_1\} & \,\, \{\Psi_1,\Psi_2\} & \,\, \cdots & \,\, \{\Psi_1,\Psi_{M_1}\} \\
\vdots & \,\, \vdots & \,\, \ddots & \,\, \vdots \\
\{\Psi_{M_1},\Psi_1\} & \,\, \{\Psi_{M_1},\Psi_2\} & \,\, \cdots & \,\, \{\Psi_{M_1},\Psi_{M_1}\} \\
\{\chi_1,\Psi_1\} & \,\, \{\chi_1,\Psi_2\} & \,\, \cdots & \,\, \{\chi_1,\Psi_{M_1}\} \\
\vdots & \,\, \vdots & \,\, \ddots & \,\, \vdots \\
\{\chi_{M_1},\Psi_1\} & \,\, \{\chi_{M_1},\Psi_2\} & \,\, \cdots & \,\, \{\chi_{M_1},\Psi_{M_1}\} \\
\end{array}
\right).\label{Delta}
\end{align}
This is a $2M_1\times M_1$ matrix and, for the Maxwell-Proca theory, the rank equals the number of Proca fields under consideration: $R=M$.
Note that $R<M_1$, which reflects the singular nature of the theory in the Dirac-Bergmann approach.
Then, it is possible to define two sets of linear combinations of the Dirac primaries
$\{\widetilde{\Psi},
\widehat{\Psi}\}$, the first set containing $R$ elements and the second set containing $S:=M_1-R=N$ elements:
\begin{align}
\widetilde{\Psi}_{\alpha}&:=\widetilde{v}_{\alpha}{}^a\Psi_a, \qquad
\widehat{\Psi}_{\overline{\alpha}}:=\widehat{v}_{\overline{\alpha}}{}^a\Psi_a, \label{prim12}
\end{align}
where we have chosen to reuse the indices introduced in (\ref{Bianchiid}) and (\ref{primlag}) for simplicity\footnote{Again, their correspondence to the
field labels in the main text is not coincidental: $(\widetilde{\Psi}_{\alpha},\widehat{\Psi}_{\overline{\alpha}})$
are constraints on the (Proca, Maxwell) fields.\label{foonotabara}}.
Here, the parameters $\widetilde{v}_{\alpha}{}^a,\widehat{v}_{\alpha}{}^a\in\mathbb{R}$ can and
should be chosen so that the newly defined
$\widetilde{\Psi}$'s and $\widehat{\Psi}$'s are linearly independent and so that
the $\widehat{\Psi}$'s are first class (primary) constraints:
\begin{align}
\{\Psi_a,\widehat{\Psi}_{\overline{\alpha}}\}\approx0\approx\{\chi_r,\widehat{\Psi}_{\overline{\alpha}}\} \qquad \forall a,\overline{\alpha},r.
\end{align}
Clearly, the $\widetilde{\Psi}$'s are second class (primary) constraints.

The procedure to distinguish between first and second class constraints among the Dirac secondaries is similar,
albeit a little more cumbersome algebraically.
First, we define the following set of linear combinations of all the Dirac secondaries and the second class Dirac primary
constraints\footnote{It is possible to trivially extend the construction to include all Dirac primaries, but this is not
necessary for our purposes.}:
\begin{align}
\chi^\prime_a:=J_a{}^b\chi_b+K_a{}^{\alpha}\widetilde{\Psi}_{\alpha} \qquad \textrm{with } J_a{}^b, K_a{}^{\alpha}\in\mathbb{R},\label{linsecdir}
\end{align}
where the parameters $J_a{}^b$ can and should be chosen so that $J$ is a non-singular matrix: $\textrm{det}(J)\neq0$.
Then, in analogy to (\ref{prim12}), it is possible to form two sets of linear combinations of the $\chi^\prime$'s:
\begin{align}
\widetilde{\chi}_{\alpha}&:=\widetilde{w}_{\alpha}{}^a\chi^\prime_a, \qquad
\widehat{\chi}_{\overline{\alpha}}:=\widehat{w}_{\overline{\alpha}}{}^a\chi^\prime_a. \label{sec12}
\end{align}
Once again, the reuse of the $(\alpha,\overline{\alpha})$ is due to the criterion explained in footnote~\ref{footalph}.
We denote by $m_2(\leq M_2)$ the number of $\widehat{\chi}$'s.
Here, $\widetilde{w}_{\alpha}{}^a,\widehat{w}_{\overline{\alpha}}{}^a\in\mathbb{R}-\{0\}$ should be chosen so that the $\widehat{\chi}$'s
are all first class
\begin{align}
\{\Psi_a,\widehat{\chi}_{\overline{\alpha}}\}\approx0\approx\{\chi_a,\widehat{\chi}_{\overline{\alpha}}\} \qquad \forall a,\overline{\alpha}, \label{allfirst}
\end{align}
for a maximal number $m_2$.
Then, the $M_2-m_2$ Dirac secondaries $\widetilde{\chi}$'s are second class constraints.
For the Maxwell-Proca theory of our interest, $m_2=N$.
This completes the derivation of table \ref{table2}.

As a double-check, we verify that (\ref{overlineDelta}), the matrix $\overline{\Delta}$ of the Poisson brackets of all the constraints that we identified as second class,
is non-singular: $\textrm{det}\overline{\Delta}\neq0$.
Note that $\overline{\Delta}$ is always an antisymmetric square matrix of dimension $R+M_2-m_2(=2M$, for the Maxwell-Proca theory).

For completeness, we indicate the relation between the first and second class constraints and the Lagrangian picture next.
As shown in the original works~\cite{DiracBergmann,AndBergmann}, linear combinations of the second class constraints $(\widetilde{\Psi},\widetilde{\chi})$
generate the Lagrangian primary (\ref{primlag}) and secondary (\ref{seclag}) constraints.
This happens through their time derivatives and within the suitable Hamiltonian subspaces, in agreement with (\ref{primprim})-(\ref{secsec}).
Observe that these constraints are all associated to the Proca fields.
Similarly, linear combinations of the first class constraints $(\widehat{\Psi},\widehat{\chi})$ generate the Bianchi identities (\ref{Bianchiid}),
all of which stem from the Maxwell fields.
Surprisingly enough, the connection between first class constraints and gauge identities has been a subject of great controversy for decades
and continues to receive attention.
Although the correct relation was long ago suggested~\cite{AndBergmann}, this was for a long time believed to be a matter of interpretation with
no physical implication.
The subject was again pondered over some thirty years later, for instance see~\cite{Costa:1985tt}, but remained as a majoritarily misconceived issue.
A lucid review of the origin and resolution of the main puzzlements to this respect appeared in~\cite{Pons:2004pp} (see also references within).
At last, the topic was shown to be of physical transcendence for the case of classical electrodynamics in~\cite{Pitts:2013uxa}.
The interested reader can consult this same reference for a meticulous historical review.
In conclusion, only a suitably tuned linear combination between a primary and secondary first class constraints is able to generate a gauge symmetry consistently.
For the Maxwell-Proca theory, this means particular compositions of $(\widehat{\Psi},\widehat{\chi})$ yield (\ref{gaugeid}).

Finally, adding the number of first (second) class constraints coming from the Dirac primaries and secondaries, we see that $N_1=2N$ ($N_2=2M$), as anticipated.
Using (\ref{ndof}), the number of propagating degrees of freedom for the proposed Maxwell-Proca theory is that of the auxiliary theory (\ref{mp}):
\begin{align}
n_{\textrm{dof}}=4(N+M)-2N-M=2N+3M. \label{DoFfin}
\end{align}
Namely, each Maxwell field propagates two degrees of freedom and each Proca field propagates three degrees of freedom.
In other words, the interaction terms between the fields in Eq.~(\ref{eq:fulllag}) supplemented by the condition (\ref{eq:secondCER}) are well-behaved, in the sense that they do not give rise to the propagation of ghost-like degrees of freedom.

As a final remark, note that we have worked in four spacetime dimensions for definiteness.
However, all the results in this appendix can be straightforwardly generalized to arbitrary dimension $d$.
The only change would be in the dimension of the configuration space $\mathcal{C}$: $M_0=d(N+M)$.
This then would yield a degree of freedom count $n_{\textrm{dof}}=(d-2)N+(d-1)M$.

\begin{widetext}
\begin{align}
\overline{\Delta}:=\left(
\begin{array}{cccccc}
\{\widetilde{\Psi}_1,\widetilde{\Psi}_1\} &\,\,  \cdots &\,\, \{\widetilde{\Psi}_1,\widetilde{\Psi}_M\} &
\,\, \{\widetilde{\Psi}_1,\widehat{\chi}_1\} & \,\,\cdots
& \,\, \{\widetilde{\Psi}_1,\widehat{\chi}_M\}\\
\vdots & \,\, \ddots  & \,\, \vdots & \,\, \vdots & \,\, \ddots  & \,\, \vdots \\
\{\widetilde{\Psi}_M,\widetilde{\Psi}_1\} &\,\,  \cdots &\,\, \{\widetilde{\Psi}_M,\widetilde{\Psi}_M\} &
\,\, \{\widetilde{\Psi}_M,\widehat{\chi}_1\} & \,\,\cdots
& \,\, \{\widetilde{\Psi}_M,\widehat{\chi}_M\}\\
\{\widehat{\chi}_1,\widetilde{\Psi}_1\} &\,\,  \cdots &\,\, \{\widehat{\chi}_1,\widetilde{\Psi}_M\} &
\,\, \{\widehat{\chi}_1,\widehat{\chi}_1\} & \,\,\cdots
& \,\, \{\widehat{\chi}_1,\widehat{\chi}_M\}\\
\vdots & \,\, \ddots  & \,\, \vdots & \,\, \vdots & \,\, \ddots  & \,\, \vdots \\
\{\widehat{\chi}_M,\widetilde{\Psi}_1\} &\,\,  \cdots &\,\, \{\widehat{\chi}_M,\widetilde{\Psi}_M\} &
\,\, \{\widehat{\chi}_M,\widehat{\chi}_1\} & \,\,\cdots
& \,\, \{\widehat{\chi}_M,\widehat{\chi}_M\}\\
\end{array}
\right), \label{overlineDelta}
\end{align}
\end{widetext}

\end{document}